\documentclass[useAMS,usenatbib,usedcolumn]{mn2e}
\pdfoutput=1
\usepackage[figuresright]{rotating}
\usepackage{lscape}
\usepackage{graphics}
\usepackage{epsfig}
\usepackage{multirow}
\usepackage{bigdelim}
\usepackage{bigstrut}
\usepackage{amsmath}
\usepackage{amssymb}
\usepackage{lscape}
\usepackage{supertabular}
\usepackage{times}
\usepackage[T1]{fontenc}
\usepackage{aecompl}
\usepackage{subcaption}
\usepackage{afterpage}
\usepackage{verbatim}

\title[The velocity field in MOND cosmology]{The velocity field in MOND cosmology}
\author[G.~N. Candlish]{G.~N. Candlish$^{1,2}$\thanks{E-mail: gcandlish@das.uchile.cl}\\
$^{1}$Departamento de Astronom\'ia, Universidad de Chile, Casilla 36-D, Santiago, Chile\\
$^{2}$Departamento de Astronom\'ia, Universidad de Concepci\'on, Casilla 160-C, Concepci\'on, Chile}
\begin{document}

\maketitle

\begin{abstract}
The recently developed code for N-body/hydrodynamics simulations in Modified Newtonian Dynamics (MOND), known as RAyMOND, is used to investigate the consequences of MOND on structure formation in a cosmological context, with a particular focus on the velocity field. This preliminary study investigates the results obtained with the two formulations of MOND implemented in RAyMOND, as well as considering the effects of changing the choice of MOND interpolation function, and the cosmological evolution of the MOND acceleration scale. The simulations are contrived such that structure forms in a background cosmology that is similar to $\Lambda$CDM, but with a significantly lower matter content. Given this, and the fact that a fully consistent MOND cosmology is still lacking, we compare our results with a standard $\Lambda$CDM simulation, rather than observations. As well as demonstrating the effectiveness of using RAyMOND for cosmological simulations, it is shown that a significant enhancement of the velocity field is likely an unavoidable consequence of the gravitational modification implemented in MOND, and may represent a clear observational signature of such a modification. It is further suggested that such a signal may be clearest in intermediate density regions such as cluster outskirts and filaments.
\end{abstract}


\section{Introduction}
The multiple successes of the $\Lambda$CDM concordance cosmology model are well known, with several independent lines of evidence supporting the model, such as observations of the temperature fluctuations in the CMB, large-scale structure statistics, weak and strong gravitational lensing due to galaxy clusters, and so on (see e.g. \cite{WeinbergCosmo,WhiteGalEvol,planck} and references therein). For further discussion of cosmological models see \cite{kroupa1} and \cite{kroupa2,kroupa3}. At smaller scales, however, the standard model of cosmology appears to exhibit certain deficiencies when confronted with observations \citep{WeinbergProbs}, at least in so far as the predictions of $\Lambda$CDM cosmological simulations are concerned. Among the various possible ``deficiencies'' of the model are: the missing satellites problem, and its sister problem of satellite halos that are ``too big to fail'' \citep{tbtf,tbtf2}; the possible planar arrangements, and co-rotation, of satellite galaxies around their hosts \citep{plane1,plane2}; the diversity of dwarf galaxy rotation curves \citep{eagledwarfcurves}, more generally known as the ``cusp/core'' problem; and the overabundance of field galaxies in $\Lambda$CDM \citep{voidprob}.

Most studies focus on the impact of an improved treatment of the baryonic physics in cosmological simulations as a remedy for these problems (for a very recent study see \citealp{sawala}, although see also \citealp{pawlowski}), but physics beyond $\Lambda$CDM may also play a role. There has been much interest recently in such ideas, motivated by the ongoing puzzle of dark energy (see \cite{beyondlcdm} for a review). Typically a modification of gravity or some new matter content is invoked to explain the late-time acceleration of the Universe, with potential impacts for non-linear structure formation (see e.g. \citealp{structform_modgrav}). An alternative modified gravity proposal, that addresses the phenomenology of dark matter rather than dark energy, is Modified Newtonian Dynamics, or MOND \citep{milgrommondoriginal}. Originally proposed as a means to obtain a galaxy rotation curve purely from its baryonic material (i.e. without the need for dark matter) the concept has been implemented in fully relativistic theories such as Tensor-Vector-Scalar (TeVeS) theory \citep{teves}, Generalised Einstein-Aether (GEA) theory \citep{GEA} and bimetric MOND \citep{bimetricmond} amongst others. Furthermore, the striking success of MOND at small scales has prompted the consideration of exotic physics for dark matter itself \citep{ddiskdm,superfluiddm,dipolardm}. A thorough review of the MOND paradigm may be found in \cite{mondreview}.

Clearly, attempting to match the large-scale structure and cosmological successes of the standard model by replacing the cold dark matter component with a modification of gravity is an enormous challenge. The predictions of these relativistic theories for the CMB power spectrum have generally shown significant discrepancy with our observed Universe \citep{skordis}. In order to test the effects of such theories on non-linear structure formation, however, we require numerical simulations using a MOND gravity simulation code.

Such a code has been recently developed by modifying the gravitational solver of the well-known and popular AMR code RAMSES \citep{teyssier}. This MONDified RAMSES, referred to as RAyMOND \citep{raymond}, utilises an extended numerical stencil and a Full Approximation Storage scheme for the Gauss-Siedel algorithm to solve the full non-linear modified Poisson equation, proposed in \cite{BekensteinMilgrom}. The fact that this equation may be derived from an aquadratic Lagrangian (thus ensuring momentum and energy conservation) leads to the name AQUAL for this specific formulation of MOND. In addition, the RAyMOND code implements the quasi-linear formulation of MOND developed specifically with numerical applications in mind, where two linear Poisson equations are solved in succession, with a modified source term for the second equation. More details regarding the RAyMOND code may be found in \cite{raymond}, as well as basic tests of the code as applied to galaxy simulations. For an alternative implementation of MOND into the RAMSES code see the PoR code of \cite{lughausen}.

In this work the applications of the RAyMOND code will be extended to cosmology, and in particular an exploration of the differences in the behaviour of the velocity field in MOND cosmologies and the standard $\Lambda$CDM model. The results of cosmological simulations using the QUMOND and AQUAL solvers will be compared, along with the effect of modifying the MOND interpolation function, and the cosmological evolution of the MOND acceleration scale. There have been several previous studies of structure formation in a cosmological context in MOND, including \cite{nusser,knebe,llinares}. These studies have emphasised that, while the matter power spectrum at $z=0$ is generally achievable in MOND (given the right initial conditions), structure formation tends to start later and proceed more rapidly than in $\Lambda$CDM. Furthermore, it is well known that MOND cannot explain the velocity dispersions of galaxy clusters, and that some amount of hidden mass is required. Therefore, more recent studies of cosmology in MOND have attempted to marry MOND with a hot dark matter particle, in the form of a light ($11$~eV) sterile neutrino \citep{angus,angus2}. The reason for choosing a hot dark matter particle is to ensure that it has a large enough free-streaming length so as not to spoil the effectiveness of MOND at galaxy scales. Unfortunately, this model overproduces high mass halos.

For this study, rather than attempt to find a cosmological solution that passes all observational tests, the focus will instead be on one potentially observable source of differences between MOND and $\Lambda$CDM: the velocity field. A previous examination of bulk velocities in the QUMOND formulation with sterile neutrino dark matter was undertaken in \cite{katz_mond_vel}. This study is also motivated by the fact that MOND remains a surprisingly successful paradigm at galaxy scales. Examining the consequences of such a modification of gravity at all scales is of considerable interest from the point-of-view of testing MOND. Furthermore, exploring outside the parameter space of $\Lambda$CDM provides an opportunity to search for possible model degeneracy and to potentially learn more about $\Lambda$CDM itself.

This paper is organised as follows: in Section \ref{mondcosmo} the MOND concept is briefly summarised, as applied to cosmology; Section \ref{sims} discusses the important issue of initial conditions for the simulations, as well as the simulations themselves; in Section \ref{results} the results are presented, with a comparison of the standard model and MOND runs, along with a demonstration of the consequences of changing the interpolation function and the cosmological dependence of the MOND acceleration scale, all with a particular focus on the velocity field. Finally we conclude in Section \ref{conclusions}.

\section{MOND and cosmology}
\label{mondcosmo}
Cosmological simulations in MOND must make several simplifying assumptions for practical reasons. A fully consistent MOND cosmological simulation (assuming that the primordial power spectrum is unchanged in a MOND context) would require the following steps: the initial power spectrum must be generated using a numerical Boltzmann solver that applies cosmological perturbation theory to some relativistic MOND theory; the initial particle positions for the simulation must then be evolved to the starting redshift using the Zel'dovich approximation (or higher order Lagrangian perturbation theory) adjusted to account for the MOND behaviour; and finally the non-linear structure must be evolved to $z=0$ using a MOND N-body/hydrodynamics code.

Previous studies of MOND cosmology have avoided the aforementioned issues by appealing to the fact that the early Universe is expected to be within the Newtonian regime on all scales, with the scale-dependent redshift of transition to the MOND regime related to the value of the MOND acceleration scale and the spectral index of the power spectrum \citep{nusser}. For smaller acceleration scales the Universe remains Newtonian until later times, and with a standard scalar spectral index of $n = 0.96$, \emph{large} scales enter the MOND regime before small scales. This study follows the usual approach to MOND simulations by assuming a background evolution that closely matches $\Lambda$CDM, as well as a standard early-time evolution. The simulations use the AQUAL and QUMOND formulations, suitably modified to co-moving coordinates, adding an additional cosmological dependence on the MOND scale to have Newtonian behaviour at early times. Possible deformations of the AQUAL or QUMOND formulations arising from a robust consideration of the non-relativistic limit of the fully relativistic MOND theory will not be investigated, although such deformations are likely to be of considerable interest. The implementation of fully consistent MOND cosmology simulations is therefore left for future work.

\subsection{AQUAL}
The AQUAL and QUMOND formulations will now be summarised, along with their use in a cosmological context, beginning with the AQUAL formulation. The modification of gravity due to MOND can be expressed in terms of a modified Poisson equation given by
\begin{equation}
\nabla \cdot \left[ \mu \left( \frac{|\nabla \phi |}{g_0} \right) \nabla \phi \right] = 4\pi G \rho
\end{equation}
where $\phi$ is the gravitational potential, $\rho$ is the density, $g_0$ is the (present-day) MOND acceleration scale and $\mu$ is a monotonically increasing function that interpolates between the Newtonian and MOND regimes. To achieve this interpolation $\mu$ must have as limiting behaviour $\mu(x) \to 1$ when $x \gg 1$ and $\mu(x) \to x$ when $x \ll 1$. Moving to the super co-moving coordinates used in RAMSES \citep{supercomoving,teyssier} this equation becomes
\begin{equation}
\label{cosmoAQUAL}
\nabla \cdot \left[ \mu \left( \frac{|\nabla \phi |}{ ag_0} \right) \nabla \phi \right] = \frac{3}{2}\Omega_m \left( \frac{\rho}{\bar{\rho}} - 1 \right)
\end{equation}
where $a$ is the scale factor, and $\phi$ is the gravitational potential arising from the density perturbations. A useful family of interpolation functions that satisfy the limiting behaviour described above is given by
\begin{equation}
\label{mondinterp}
\mu(x) = \frac{x}{(1+x^n)^{1/n}}
\end{equation}
where $n \geq 1$. A cosmological dependence may be added by making the MOND scale a function of the scale factor. For this work, the simple prescription of \cite{llinares} is used:
\begin{equation}
g_M = \gamma(a)g_0
\end{equation}
where $\gamma(a)$ is some function of the scale factor. This function is chosen to be 
\begin{equation}
\label{gamma}
\gamma(a) = a^m,
\end{equation}
with $m = 0, 1$ or $2$. The MOND scale is therefore either a constant, or multiplied by some power of the scale factor. Note the additional appearance of the scale factor in Eq. ~\ref{cosmoAQUAL} due to the transformation to co-moving coordinates. This use of $\gamma(a)$ allows a phenomenologically motivated non-trivial cosmological behaviour for the MOND effect, which may arise from the full relativistic theory. For $m > 0$ it is clear that the MOND scale is suppressed at early times, consistent with the use of standard initial conditions for the simulations.

\subsection{QUMOND}
The QUMOND formulation recovers the MOND phenomenology by using two Poisson equations, solved consecutively. The first is the standard Newtonian equation, while the second uses a modified source term, which may be interpreted as an additional density component, often referred to as ``phantom dark matter.'' This additional density is derived from the Newtonian potential as follows:
\begin{equation}
\label{phantomdm}
\rho_{PDM} = \frac{1}{4\pi G} \nabla \cdot \left[ \tilde{\nu} \left( \frac{|\nabla \phi_N|}{g_0} \right) \nabla \phi_N \right]
\end{equation}
where $g_0$ is again the (present-day) MOND scale, and $\phi_N$ is the Newtonian gravitational potential calculated from the standard Newtonian Poisson equation. The function $\tilde{\nu} = \nu - 1$, where $\nu$ is the inverse of the interpolation function given in Eq.~\ref{mondinterp}:
\begin{equation}
\label{qumondinterp}
\nu(y) = \left( \frac{1}{2} + \frac{1}{2}\sqrt{1 + \frac{4}{y^n}} \right)^{1/n}.
\end{equation}
Transforming Eq.~\ref{phantomdm} to super co-moving coordinates this becomes
\begin{equation}
\rho_{PDM} = \frac{2}{3\Omega_m} \nabla \cdot \left[ \tilde{\nu} \left( \frac{|\nabla \phi_N|}{a g_0} \nabla \phi_N \right) \right],
\end{equation}
where $\phi_N$ is the Newontian potential arising from the density perturbations, neglecting the background. As for the AQUAL formulation, a cosmological dependence on the MOND scale is incorporated by using $g_M = \gamma(a)g_0$ for some chosen function of the scale factor $\gamma(a)$, which is again taken to be Eq.~\ref{gamma} with $m=0,1$ or $2$. Note that the ``phantom dark matter'' density may take negative values.

In all simulations $g_0 = 1.2 \times 10^{-10}$m s$^{-2}$.

\section{Simulations}
\label{sims}

\subsection{Initial conditions}
\label{ics}
In addition to the possible implications of relativistic MOND on the primordial perturbations, and the consequent impacts on observables such as the CMB, it is important to address the question of the initial conditions required for simulations of cosmological structure formation well into the non-linear regime. As stated earlier, the background cosmology will be assumed to be that given by General Relativity, with $H_0 = 70$~(km/s)/Mpc, as well as standard Gaussian primordial perturbations with a spectral index of $n=0.96$. Where the models in this study will depart from $\Lambda$CDM is in the choice of matter/energy content.

From the point-of-view of the original motivation of MOND, it may seem appealing to neglect dark matter entirely, and only include baryons. This approach will fail, at least when using a standard GR cosmology at early times, as it is well known that the baryon-photon coupling before recombination results in a huge damping of structure on small scales, such that by $z=0$ there is no time for small-scale structure to form, even in a MOND context. Therefore a small amount of collisionless dark matter will be included in the simulations, and the baryonic material will be neglected (i.e. these are collisionless N-body simulations without hydrodynamics). This approach may be partly justified by the expectation that, even in a MOND Universe, there is some non-baryonic dark matter. This is because galaxy clusters typically have accelerations beyond the MOND regime \citep{mondgalclusters}, and thus any ``missing matter'' cannot be accounted for by a modification of gravity, but only by the presence of additional mass, often taken to be in the form of a sterile neutrino \citep{sterileneutrinos}. The observations of the CMB also strongly support the existence of additional non-baryonic matter. While some future relativistic MOND theory may allow for initial conditions without any additional collisionless matter, such questions are beyond the scope of this study. In addition, the inclusion of baryons in the simulations implies the inclusion of sub-grid physics recipes with many adjustable parameters, and increased computational time. To avoid these complications, these preliminary simulations neglect the baryons.

\begin{figure}
\centering
\includegraphics[width=7.0cm]{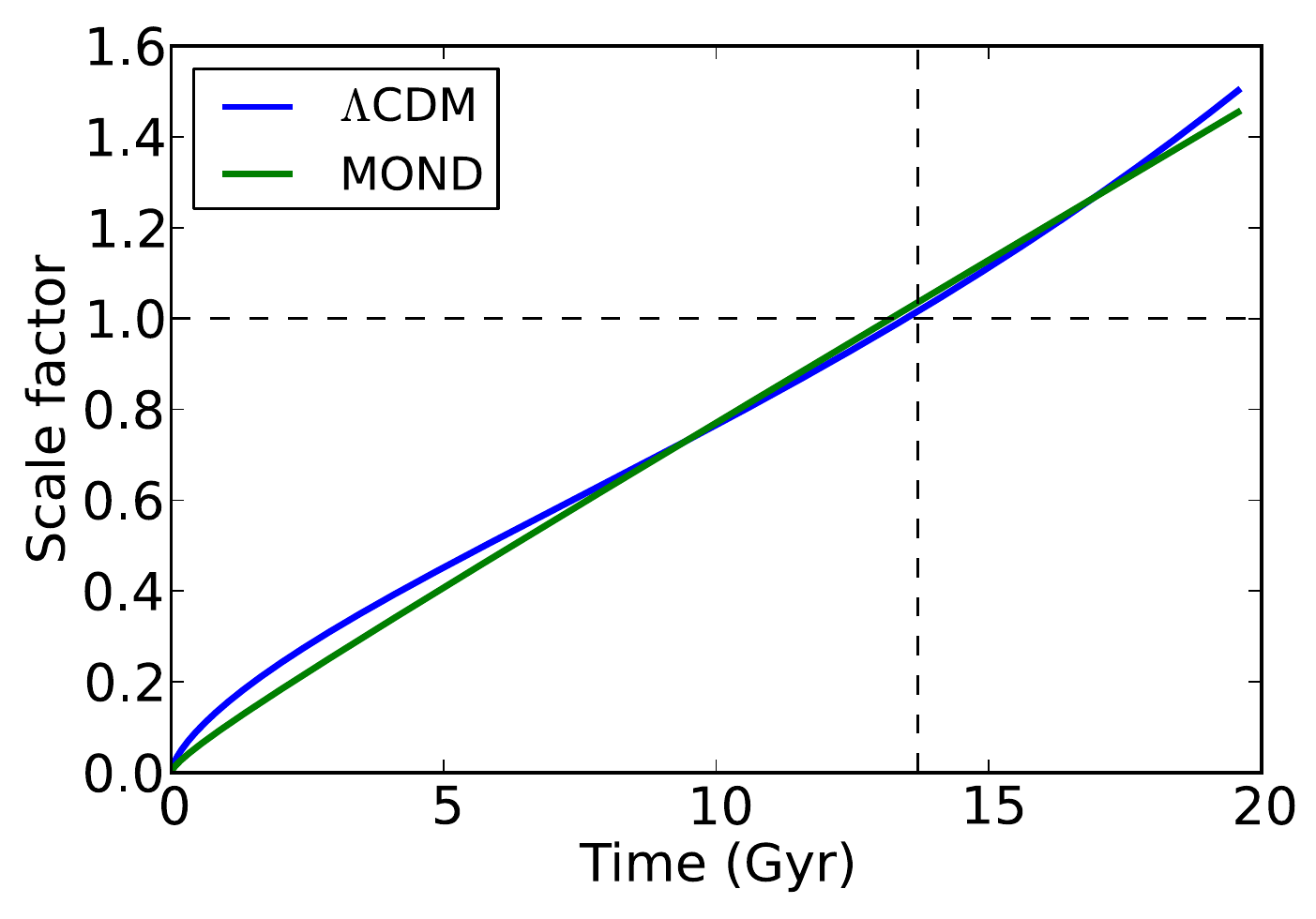}
\caption{Evolution of the scale factor with time, using the parameters of the $\Lambda$CDM model and the MOND models. The dashed black lines indicate $a=1$ and $t=13.7$~Gyr.}
\label{bgndevolution}
\end{figure}

Due to the lower matter content in the MOND simulations, the dark energy component must be removed completely (i.e. $\Omega_{\Lambda} = 0$ in all the MOND simulations) to ensure that the background expansion is approximately consistent with that of $\Lambda$CDM, as the MOND models will then have a large contribution from the curvature component to the background evolution. This is illustrated in Fig.~\ref{bgndevolution} where the time evolution of the scale factor is shown for the parameters used in our $\Lambda$CDM model compared with those used for the MOND models. The vertical dashed black line is set at $t=13.7$~Gyr, while the horizontal dashed black line is at $a=1$. Clearly the background evolution is very similar in both cases, and so any differences in structure formation are almost entirely due to the modification of gravitational clustering in MOND, and not because of a modified expansion history. One may consider these choices of parameters as those required to ensure an ``effective'' MOND cosmology, given the absence of a fully consistent MOND cosmological model.

\begin{figure}
\centering
\includegraphics[width=7.0cm]{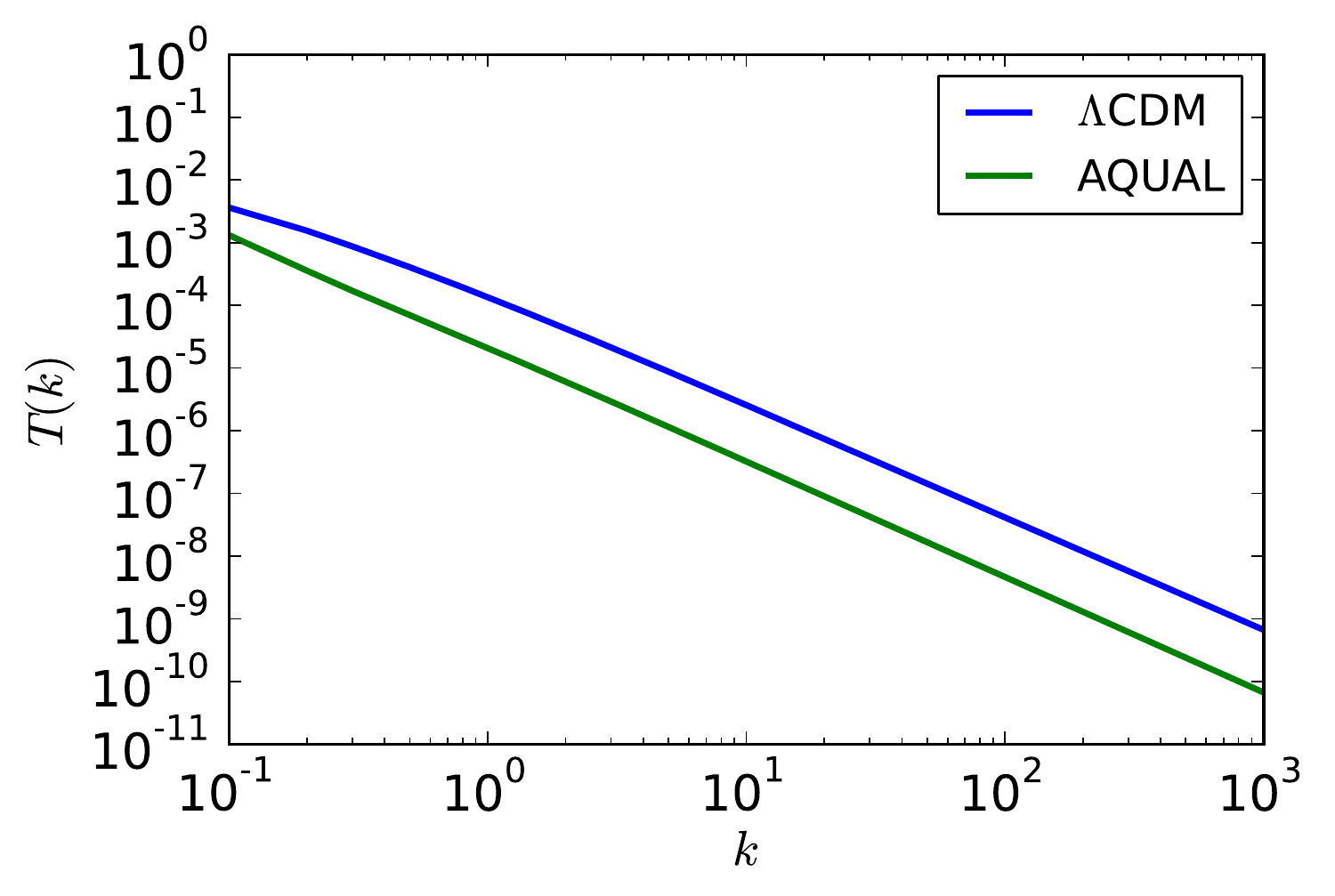}
\caption{Transfer functions used in the simulations in this paper, initially generated by CAMB, and then normalised using $\sigma_8$ by MUSIC.}
\label{transferfunctions}
\end{figure}

The transfer functions used to determine the initial power spectra are generated using CAMB \citep{cambpaper} for the matter content appropriate to each simulation. These transfer functions are then used to generate the initial conditions at $z=50$ using the MUSIC code \citep{music}. The normalisation of the power spectrum in MUSIC for the MOND simulations is set to $\sigma_8 = 0.25$, while the $\Lambda$CDM simulation uses $\sigma_8 = 0.88$. It should be noted that there is no physical content to the use of $\sigma_8$ for the MOND simulations, as this parameter assumes a linear power spectrum evolved using a standard cosmological model. The resulting normalised transfer functions used in MUSIC are shown in Fig.~\ref{transferfunctions}. The standard $\Lambda$CDM simulation in Newtonian gravity (with typical choices for the cosmological parameters, given in Table \ref{simparamstable}) is run using the upper transfer function, while the simulations using MOND gravity with a significantly reduced matter content use the lower transfer function. The latter is clearly significantly suppressed in comparison to that of a standard CDM content. Even with this much reduced initial power spectrum, the MOND gravitational enhancement is able to form a comparable amount of small-scale structure at $z=0$, in addition to a slight excess of large-scale structure.

In summary, the rationale for these initial conditions is that the MOND simulations reproduce reasonably well the amount of structure formed in $\Lambda$CDM by $z=0$ (as measured by the power spectrum), making the density field an \emph{a posteriori} fitted quantity. The velocity fields are then studied for possible observational signatures of MOND.

\subsection{Summary of simulations}
Firstly, the differences between a standard $\Lambda$CDM simulation and the two MOND formulations included in the RAyMOND code, AQUAL and QUMOND, are studied. These simulations use the so-called ``simple'' MOND interpolation function, given by Eq.~\ref{mondinterp} with $n=1$. In addition, the effects of a more rapid transition between the Newtonian and MONDian regimes are studied by using Eq.~\ref{mondinterp} with $n=5$ in a QUMOND run. These simulations will be referred to as $\Lambda$CDM (or the standard model), AQUAL, QUMOND and QUMONDn5 respectively. All of these simulations use $m=1$ in Eq.~\ref{gamma}. To test the effects of changing the cosmological dependence of the MOND scale, two  additional AQUAL simulations are run, one where $m=0$ in Eq.~\ref{gamma} (i.e. no cosmological dependence), and one where $m=2$ (i.e. a rapid cosmological transition). These simulations are referred to as AQUALnoCD and AQUALfastCD respectively. The removal of any cosmological dependence for the MOND scale in the AQUALnoCD run violates the assumption that the Universe is Newtonian at the time of the initial conditions, but the results of this simulation are nevertheless illuminating.

All simulations use a co-moving box size of $32 h^{-1}$~Mpc, with $128^3$ particles. Although this is a small box size, it is sufficient for the purposes of comparison between the models. For the $\Lambda$CDM simulations the particle mass is $1.6 \times 10^9 M_{\odot}$, while for the MOND simulations, due to the reduced matter content, the particle mass is $2.5 \times 10^8 M_{\odot}$. All the simulation parameters are summarised in Table~\ref{simparamstable}.

\begin{table*}
\begin{center}
\begin{tabular}{ccccccccc}
\hline
Simulation & $m$ & $n$ & Box size ($h^{-1}$~Mpc) & Particles & $\Omega_{CDM}$ & $\Omega_{b}$ & $\Omega_{\Lambda}$ & $\sigma_8$ \\
\hline
$\Lambda$CDM & - & - & $32$ & $128^3$ & $0.26$ & $0.04$ & $0.7$ & $0.88$ \\
AQUAL       & $1$ & $1$ & $32$ & $128^3$ & $0.04$ & $0.04$ & $0$ & $0.25$ \\
QUMOND      & $1$ & $1$ & $32$ & $128^3$ & $0.04$ & $0.04$ & $0$ & $0.25$ \\
QUMONDn5    & $1$ & $5$ & $32$ & $128^3$ & $0.04$ & $0.04$ & $0$ & $0.25$ \\
AQUALnoCD   & $0$ & $1$ & $32$ & $128^3$ & $0.04$ & $0.04$ & $0$ & $0.25$ \\
AQUALfastCD & $2$ & $1$ & $32$ & $128^3$ & $0.04$ & $0.04$ & $0$ & $0.25$ \\
\end{tabular}
\end{center}
\caption{Parameters used in all the simulations considered in this work. The integers $m$ and $n$ are the powers used in the cosmological dependence of the MOND scale and the MOND interpolation function (Eqs.~\ref{gamma} and \ref{mondinterp} respectively). The chosen gravitational theory is indicated by the simulation name. The transfer functions are generated by CAMB using a standard spectral index of $n=0.96$ for the Gaussian perturbations, and a standard background FLRW evolution with $H_0 = 70$~(km/s)/Mpc.}
\label{simparamstable}
\end{table*}

The additional complications of a non-linear solver with extended numerical stencil means that the RAyMOND code is somewhat slower than standard RAMSES, as discussed in \cite{raymond}. For this work the manner in which the code checks for numerical convergence has also been modified: the code checks that the norm of the residual is sufficiently small (the specific value chosen is $10^{-9}$) before declaring convergence. This sometimes causes the solver to take significantly longer to declare convergence than before, but ensures that the numerical solution is acceptably accurate.

\section{Results}
\label{results}
\subsection{Density fields}

\begin{figure*}
\centering
\includegraphics[width=14.0cm]{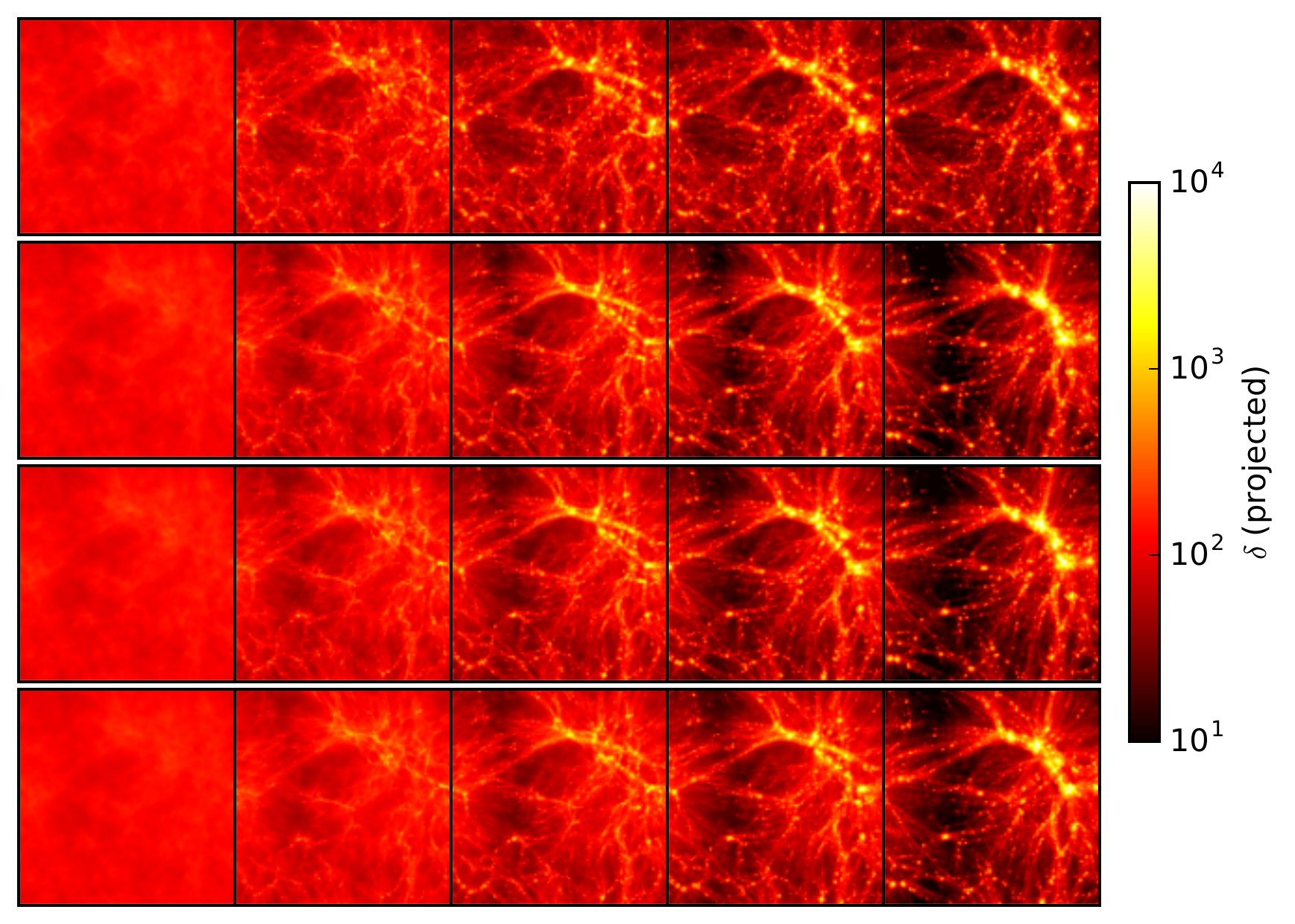}
\caption{Projection plots (along the $z$ axis) of the full simulation volume, showing the evolution of the density field, extracted to a regular grid using the DTFE utility. These projections are simple integrals (sums) along the $z$ axis at each point in the $x-y$ plane. The scale factors of each panel from left to right are $0.1, 0.3, 0.5, 0.7$ and $1.0$, corresponding to redshifts $9.0, 2.3, 1.0, 0.4$ and $0$. The top row of panels is the $\Lambda$CDM model, the second row is the AQUAL run, the third row is the QUMOND run, and the last row is the QUMONDn5 run.}
\label{densityplots}
\end{figure*}

\begin{figure*}
\centering
\begin{tabular}{cc}
\includegraphics[width=7.0cm]{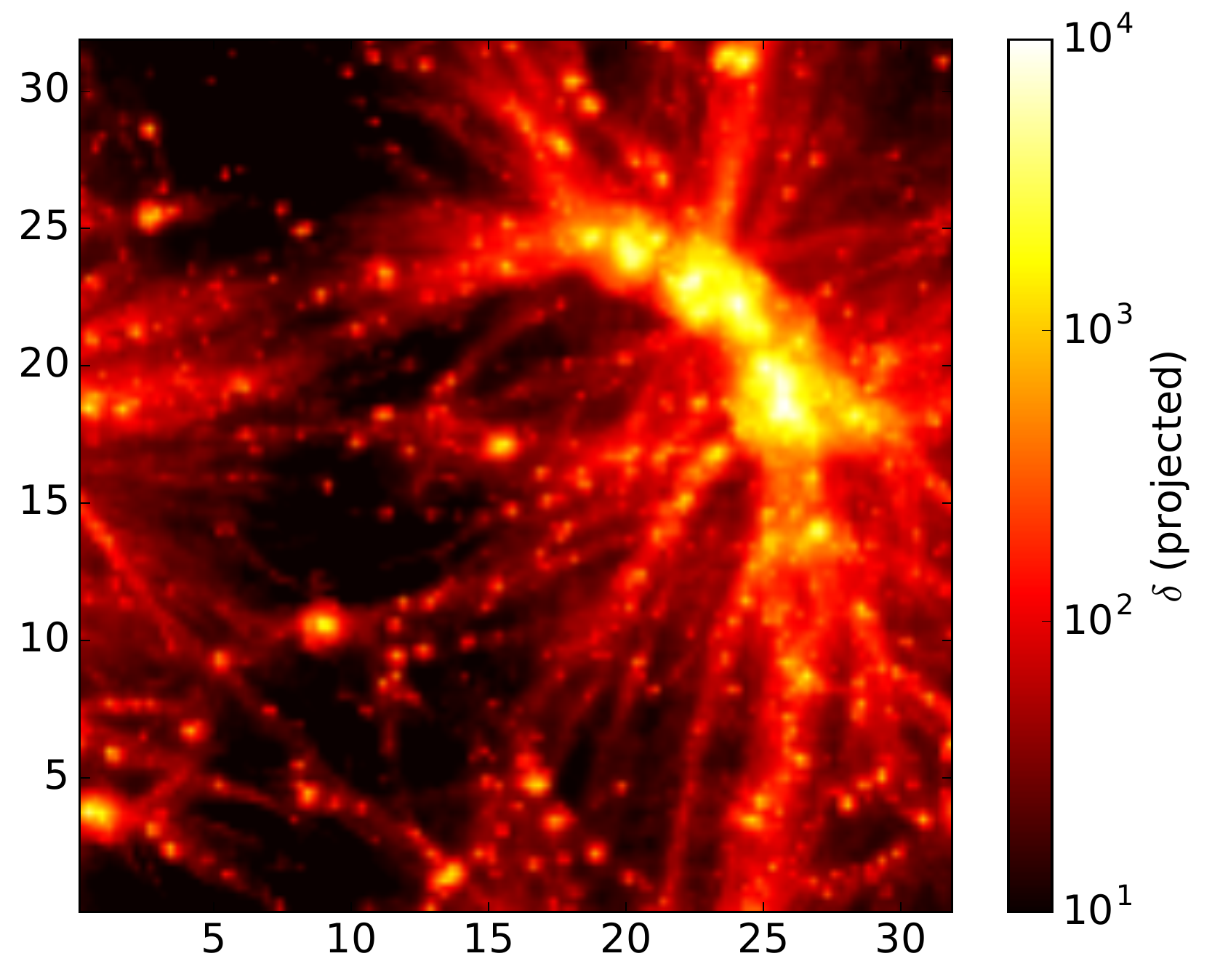} & \includegraphics[width=7.0cm]{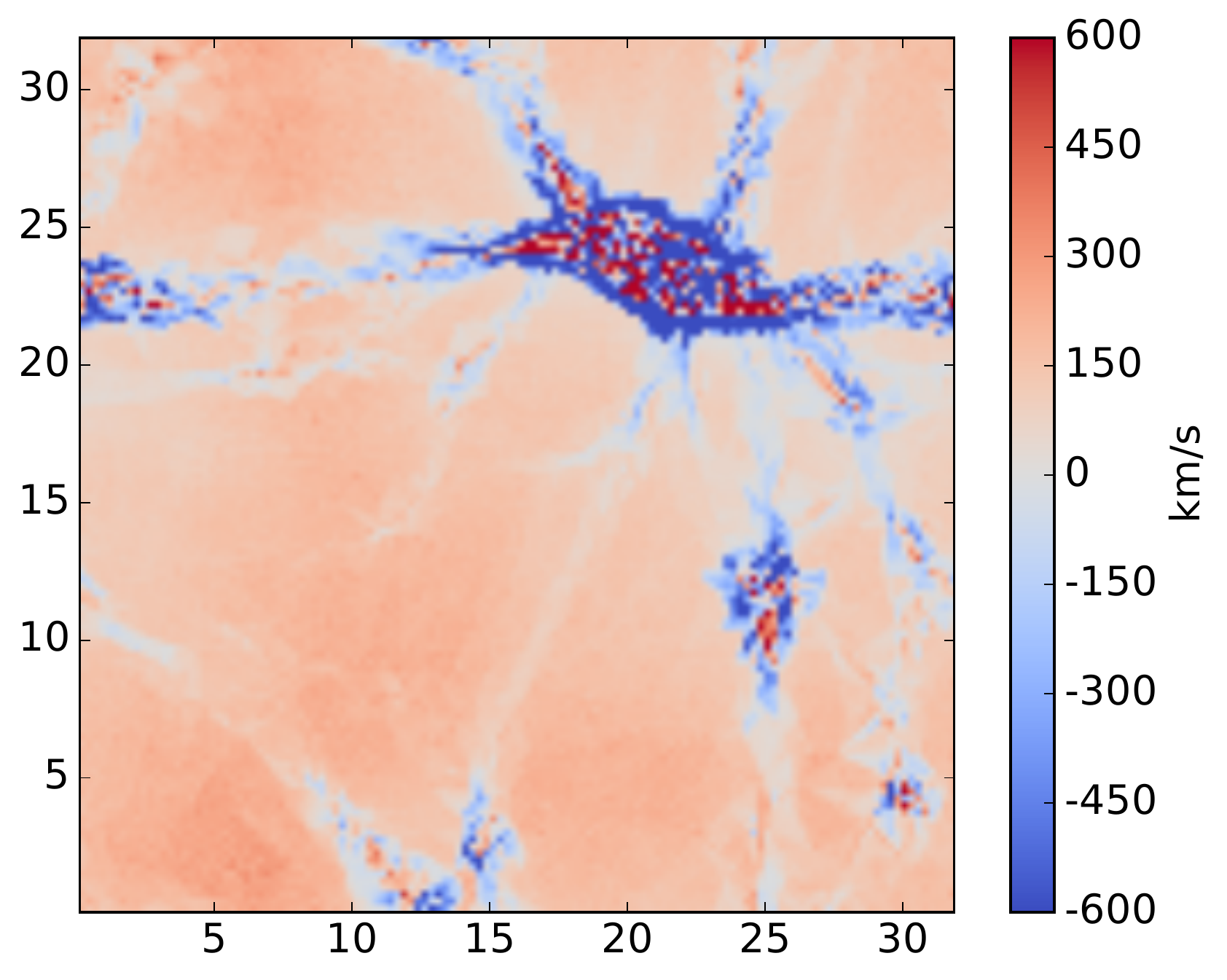}
\end{tabular}
\caption{Projected matter density field and velocity divergence field (slice) for the AQUALnoCD model, at $a=0.3$.}
\label{noCosmoDep_densityVelDiv}
\end{figure*}

\begin{figure*}
\centering
\begin{tabular}{cc}
\includegraphics[width=7.0cm]{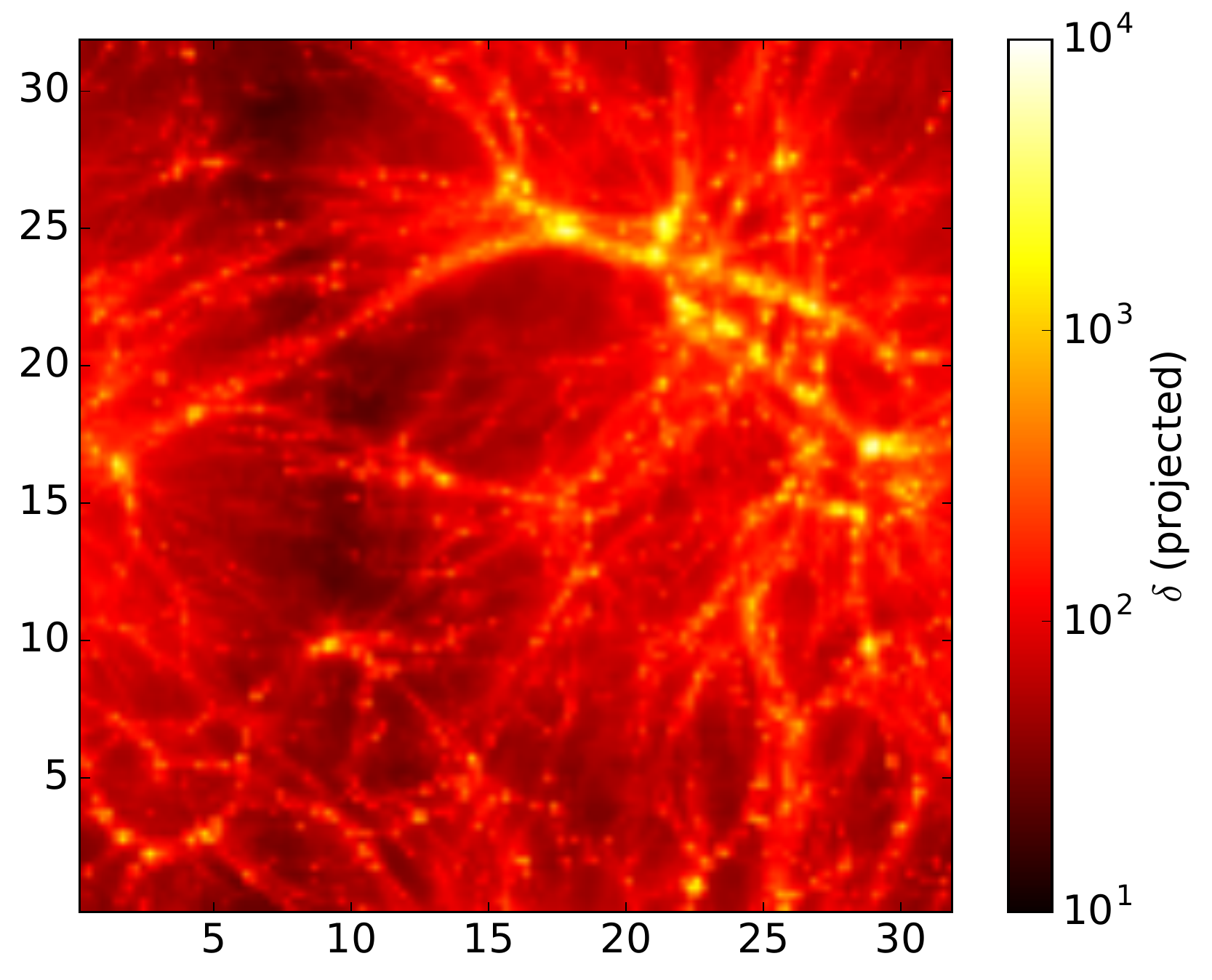} & \includegraphics[width=7.0cm]{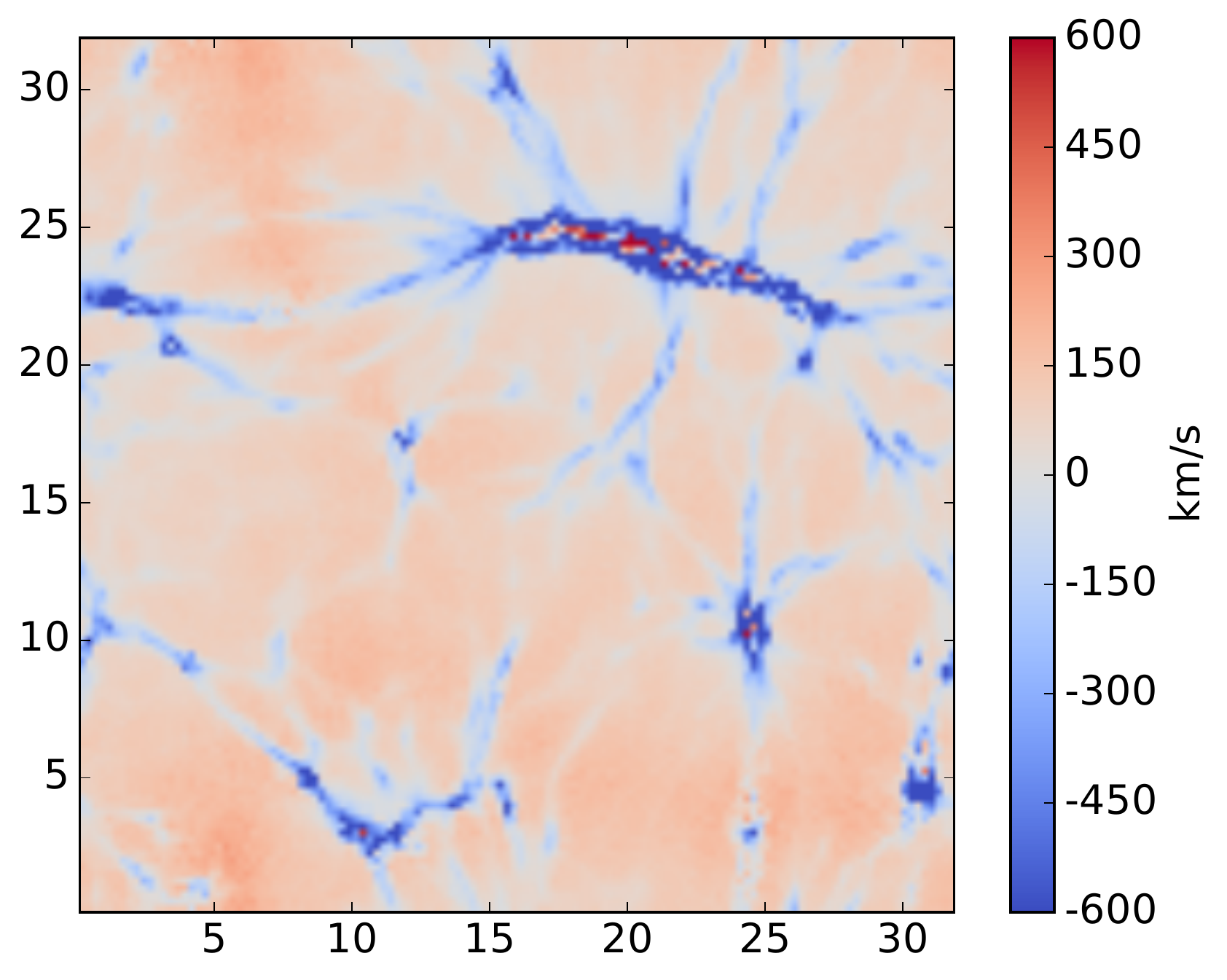}
\end{tabular}
\caption{Projected matter density field and velocity divergence field (slice) for the AQUALfastCD model, at $a=1.0$.}
\label{fastCosmoDep_densityVelDiv}
\end{figure*}

In order to construct density and velocity fields from the simulations the Delauny Tesselation Field Estimator (DTFE) technique \citep{dtfe1,dtfe2,dtfe3} is used. This allows for the extraction of these fields from the particle locations to regular grids.

The density distributions in each of the simulations at a selection of redshifts are shown in Fig.~\ref{densityplots}, where the density is projected into the plane along the $z$-axis of the simulation volume. While the precise density distribution differs between the $\Lambda$CDM and MOND simulations, the overall amount of structure is broadly similar by $z=0$. This is very much not the case for the AQUALnoCD model, whose density field is shown in the left panel of Fig.~\ref{noCosmoDep_densityVelDiv} at the snapshot $a=0.3$. Even at this very early time in the simulation, this model has already formed a great deal of structure. At the opposite extreme, in the left panel of Fig.~\ref{fastCosmoDep_densityVelDiv} (at snapshot $a=1$) we can see that the AQUALfastCD model, with a suppressed MOND effect at earlier times and the same reduced matter content as the other models, leads to a less clustered density field at late times. We can also see that the QUMONDn5 model, with a more rapid interpolation from MONDian gravity to Newtonian, shows marginally less clustering by the end of the simulation.

It is important to recall that the initial conditions have been specifically chosen to match the amount of structure at $z=0$ for the AQUAL, QUMOND and $\Lambda$CDM simulations\footnote{The same initial conditions have been used for the other MOND models as well, therefore those models are \emph{not} tuned to match the $\Lambda$CDM power spectrum at $z=0$.}. A clear distinction between the MOND simulations and the standard model, and one that has been discussed in previous studies of structure formation in MOND, is the reduced amount of structure at early times in the MOND simulations. This is more clearly seen in the evolution of the matter power spectrum, shown in Fig.~\ref{powspectime} and discussed in Section~\ref{section:powspectra}.

This evolution of structure should be interpreted within the confines of the assumptions made in this study: i.e. that non-relativistic standard MOND is applicable at the very low (in cosmological terms) redshifts of $z < 50$. As previously stated, without some significant deviation from this standard MOND behaviour, structure formation must proceed in a MOND Universe according to the results of our simulations, given our chosen initial conditions.

\begin{figure}
\centering
\includegraphics[width=7.0cm]{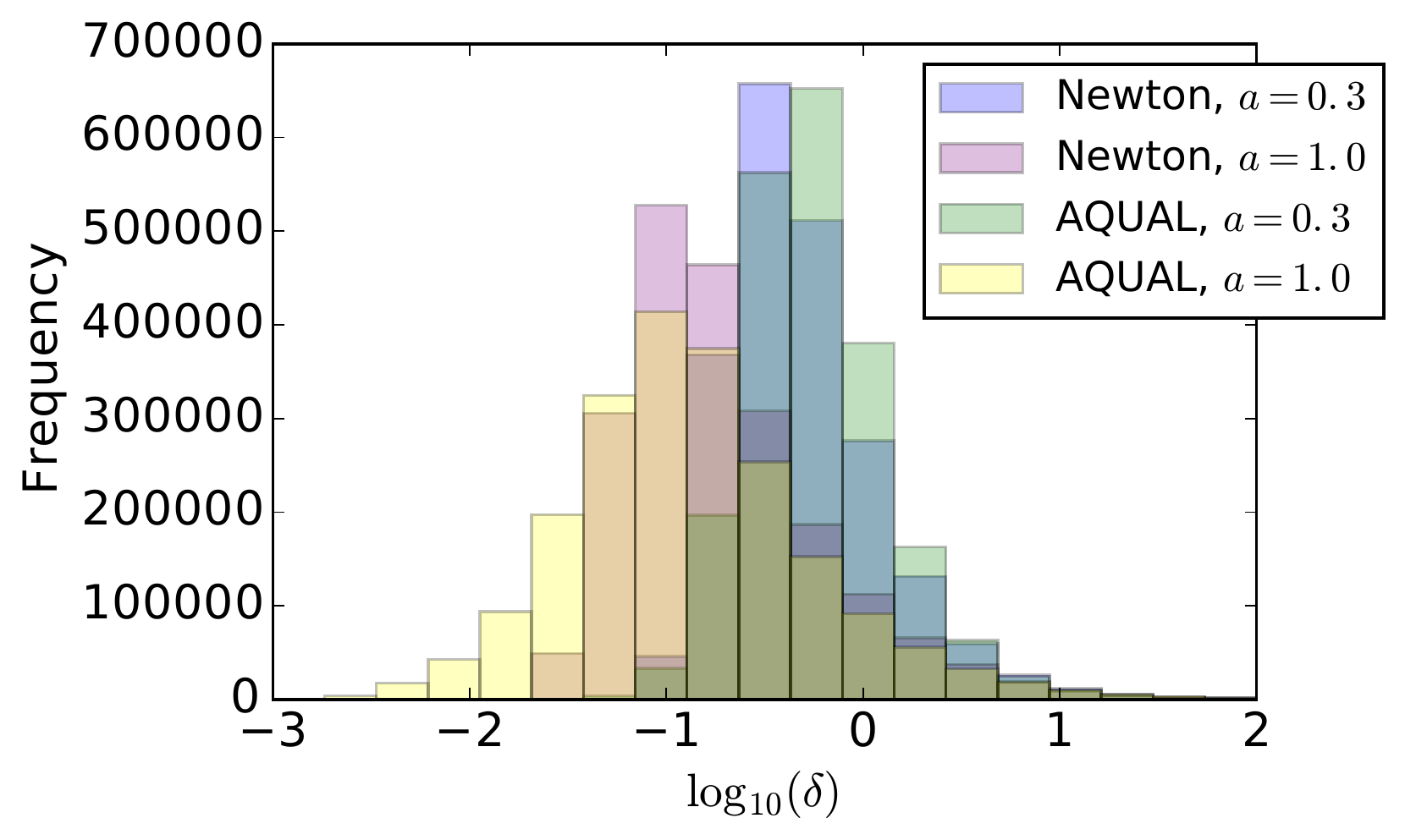}
\caption{Frequency distribution of density values from the DTFE field, at $a=0.3$ (blue and green for $\Lambda$CDM and AQUAL respectively) and $a=1.0$ (purple and yellow for $\Lambda$CDM and AQUAL respectively).}
\label{densityDist}
\end{figure}

The frequency distribution of the density fields on the regular grid extracted by DTFE is shown in Fig.~\ref{densityDist}, with density bins in $\log$-space, for the $\Lambda$CDM and AQUAL runs at snapshots $a=0.3$ and $a=1$. As the density field clusters over time, the number of (fixed) grid points with large overdensities becomes smaller. Thus this plot emphasises the growth of the \emph{underdensities} in both models, as the densities at a larger number of grid points fall well below the normalised average value of $1$. Thus the distributions evolve to the left, as the underdense regions increase in size. At early times, the clustering is slightly more evolved in the $\Lambda$CDM run than in AQUAL, as shown by the standard model distribution being peaked slightly to the left of the AQUAL distribution. This is reversed by late times, with the AQUAL distribution showing a more extended tail of underdense regions.

\subsection{Velocity divergence fields}

\begin{figure*}
\centering
\includegraphics[width=14.0cm]{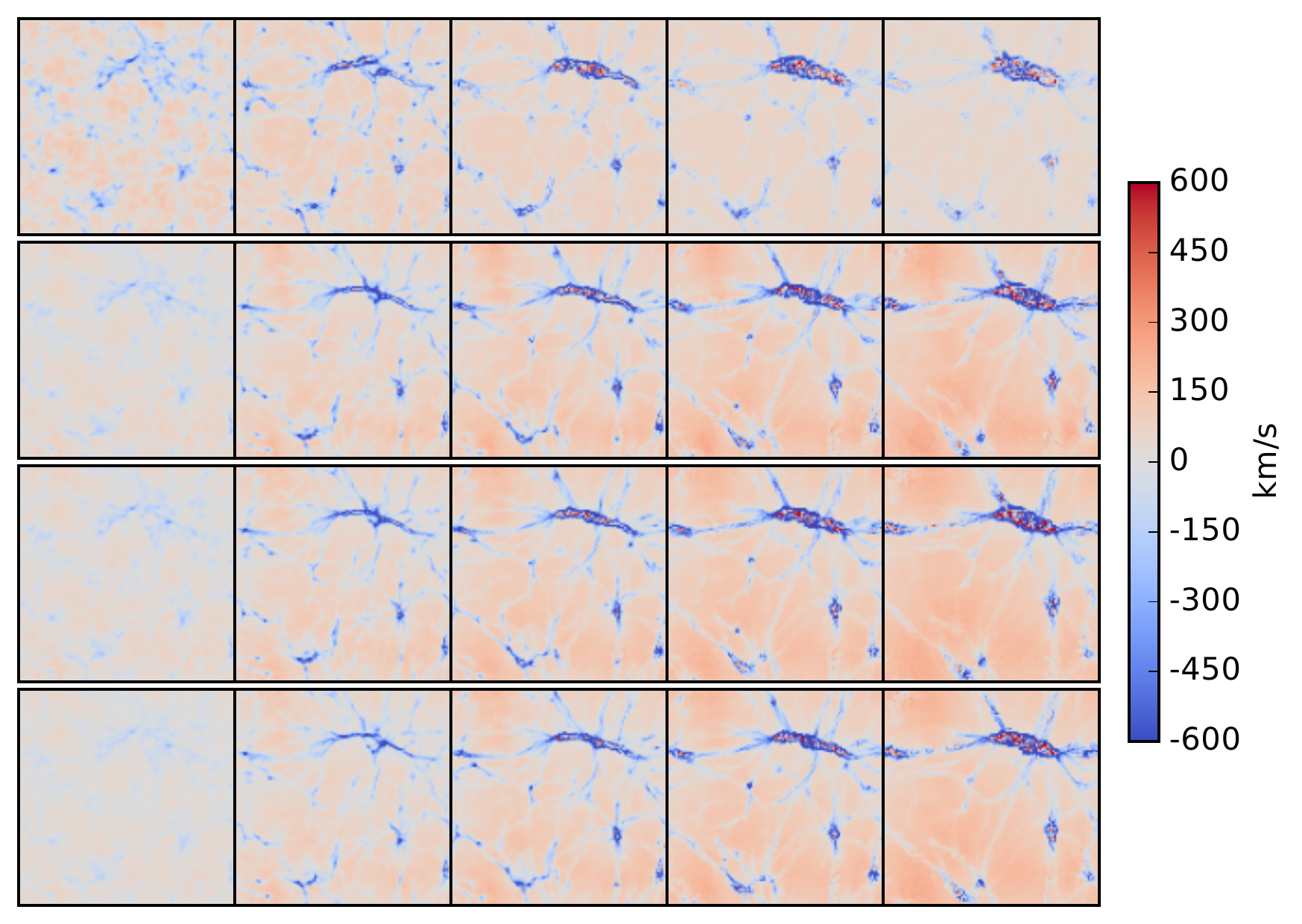}
\caption{Velocity divergence field for a slice approximately half-way along the $z$ axis of the box, determined from the velocity field on a regular grid using the DTFE code. The scale factors of each panel from left to right are $0.1, 0.3, 0.5, 0.7$ and $1.0$, corresponding to redshifts $9.0, 2.3, 1.0, 0.4$ and $0$. The top row of panels is the $\Lambda$CDM model, the second row is the QUMOND run, the third row is the AQUAL run, and the last row is a QUMOND run with $n=5$ in the MOND interpolation function.}
\label{veldivplots}
\end{figure*}

The velocity divergence fields extracted by the DTFE code are shown in Fig.~\ref{veldivplots}. These plots show a slice through the full volume, at approximately the mid-way point of the $z$-axis. At high redshift, due to the suppression of the MOND effect, and the reduced matter content, the velocity divergence in the MOND simulations is slightly reduced compared to that of the Newtonian simulations. By $z=1$, however, the velocity divergence is more pronounced in the MOND simulations, especially in underdense regions. In fact, it appears that the evolution of the velocity divergence field in underdense regions is opposite in the two cases: in the $\Lambda$CDM model the divergence is strongest at early times, but begins to diminish once the structures have formed and the voids ``stabilise.'' In the MOND simulations, the velocity divergence continues to increase, right up to $z=0$, illustrating the MOND effect driving matter out of the void regions and into structures. In the MOND simulations we can also see higher values of the velocity divergence in regions that are feeding the large structure forming in the upper right of the plot, as well as more sharply pronounced infall of material into the filaments. This may suggest that one should look for differences in the velocity field either in underdense void regions, or in the moderate density regions that are feeding into virialised structures.

The AQUALnoCD and AQUALfastCD velocity divergence fields are shown in the right panels of Figs.~\ref{noCosmoDep_densityVelDiv} and \ref{fastCosmoDep_densityVelDiv} respectively. The rapid formation of structure leads to strong velocity divergences in the AQUALnoCD run at early times ($a=0.3$), with obvious clearing of void regions and the feeding of large overdensities. In fact, even by this early time in the simulation, the overdensity forming in the upper right of the chosen slice is more developed than in the $z=0$ plots of the other MOND models. The AQUALfastCD run, on the other hand, has a somewhat smoother velocity divergence field at late times ($a=1$), with less pronounced void clearance, and the preservation of some filamentary structure. Comparing with Fig.~\ref{veldivplots}, we can see that this filamentary structure corresponds more to an earlier stage of the other MOND simulations, or the $\Lambda$CDM model, consistent with the delayed onset of structure formation in this model.

\subsection{Power spectra}
\label{section:powspectra}
\begin{figure}
\centering
\includegraphics[width=7.0cm]{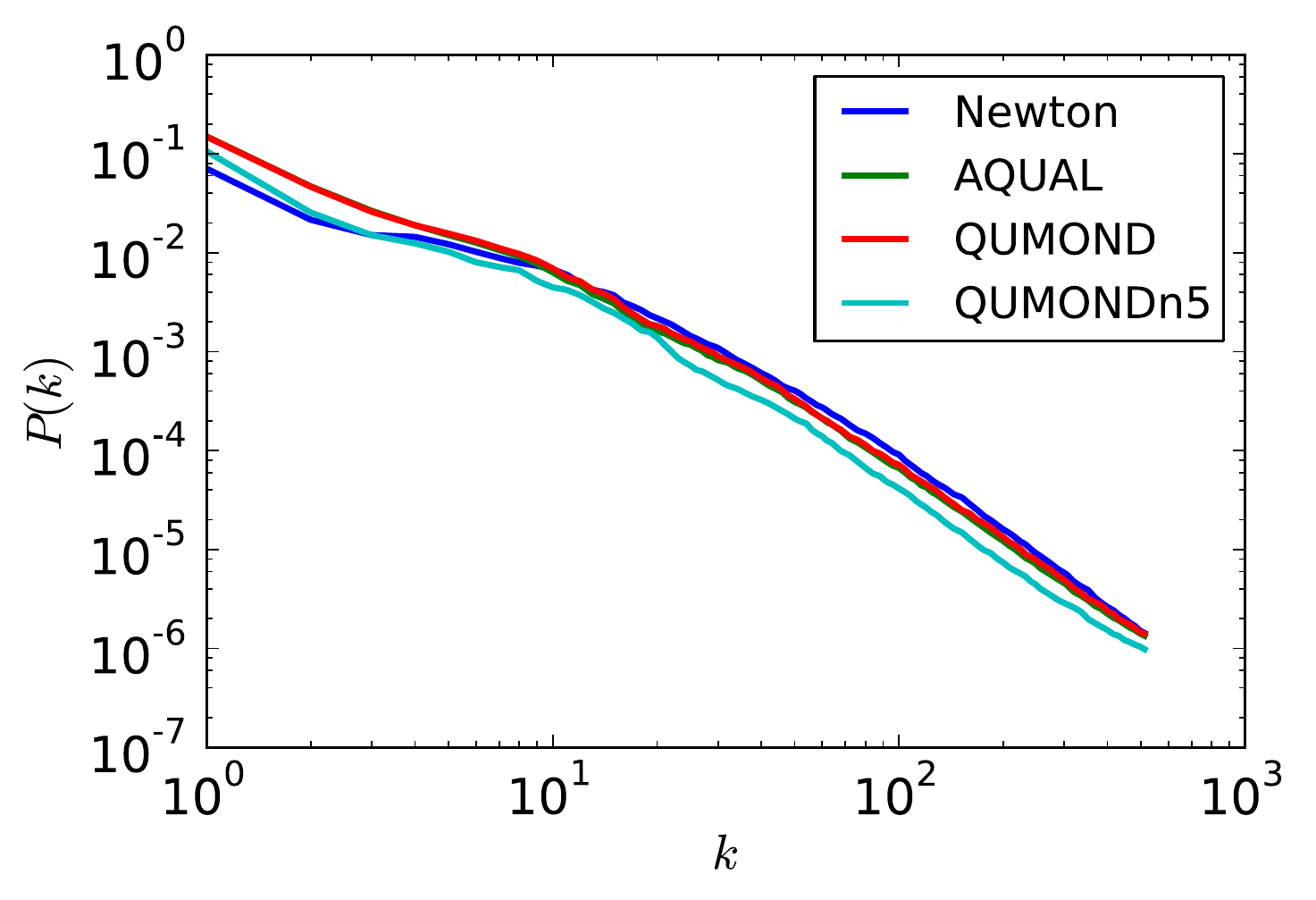}
\caption{Power spectra for each model at $z=0$, calculated using POWMES.}
\label{powerspec}
\end{figure}

As a more quantitative check of the amount of structure formation, the $z=0$ matter power spectrum is shown in Fig.~\ref{powerspec}, calculated using the POWMES utility \citep{powmes}. It is clear that the overall structure formation in the MOND and $\Lambda$CDM simulations is similar, especially at smaller scales, as ensured by tuning the initial conditions for the AQUAL and QUMOND simulations. At large scales, however, there is a difference between the MOND models and $\Lambda$CDM. As remarked in \cite{nusser}, the counter-intuitive situation can arise in MOND where the large scale perturbations enter the MOND regime \emph{before} the smaller scale perturbations, assuming a spectral index for the initial perturbations that satisfies $n > -1$ (as in these models). As such, the long-wavelength perturbations spend more time in a regime of enhanced gravity, leading to more large-scale power than in the Newtonian case.

Note that the QUMONDn5 model has not been tuned in order to closely match the $z=0$ power spectrum of $\Lambda$CDM. Thus we can compare the power spectrum in this model to see the effect of a more rapid transition from the MOND to Newtonian regime. Clearly there is an overall suppression of structure formation at all scales. This is because, although most scales may still be within the MOND regime, for a fixed ratio of gravitational acceleration to the MOND scale, a more rapid transition function takes values closer to $1$, slightly diminishing the MOND effect across all scales.

\begin{figure*}
\centering
\begin{tabular}{ccc}
\includegraphics[width=6.0cm]{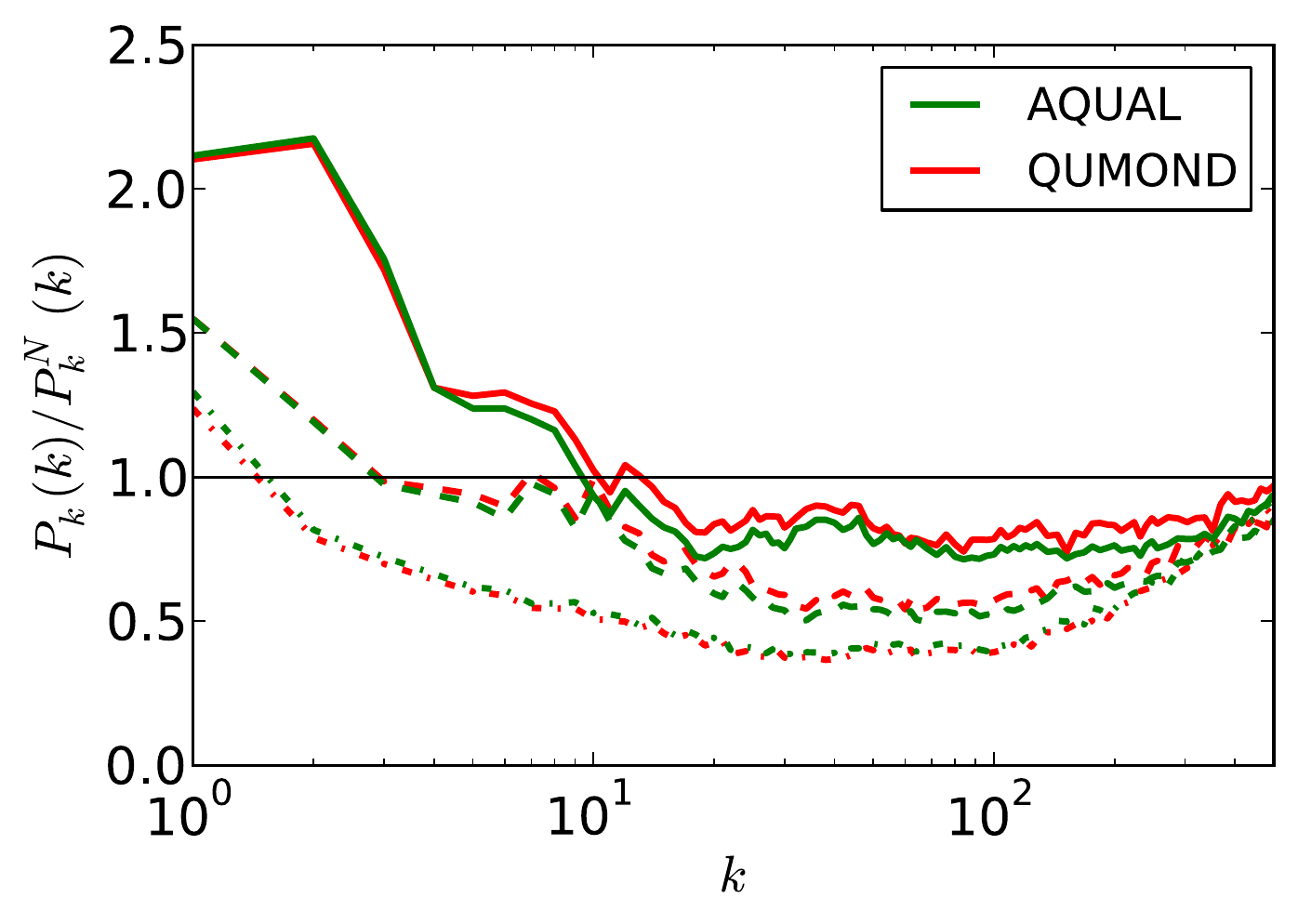} & \includegraphics[width=6.0cm]{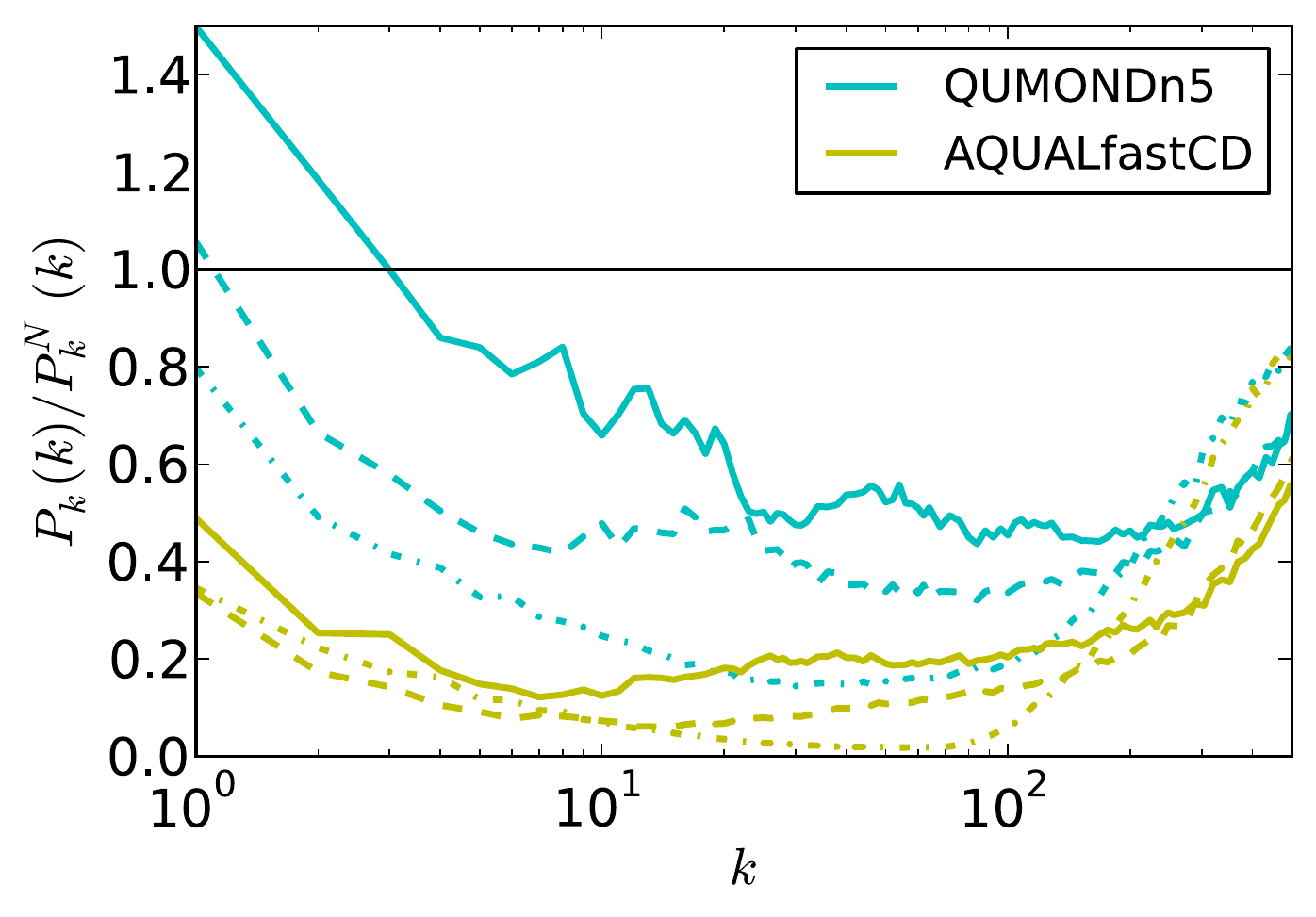} & \includegraphics[width=6.0cm]{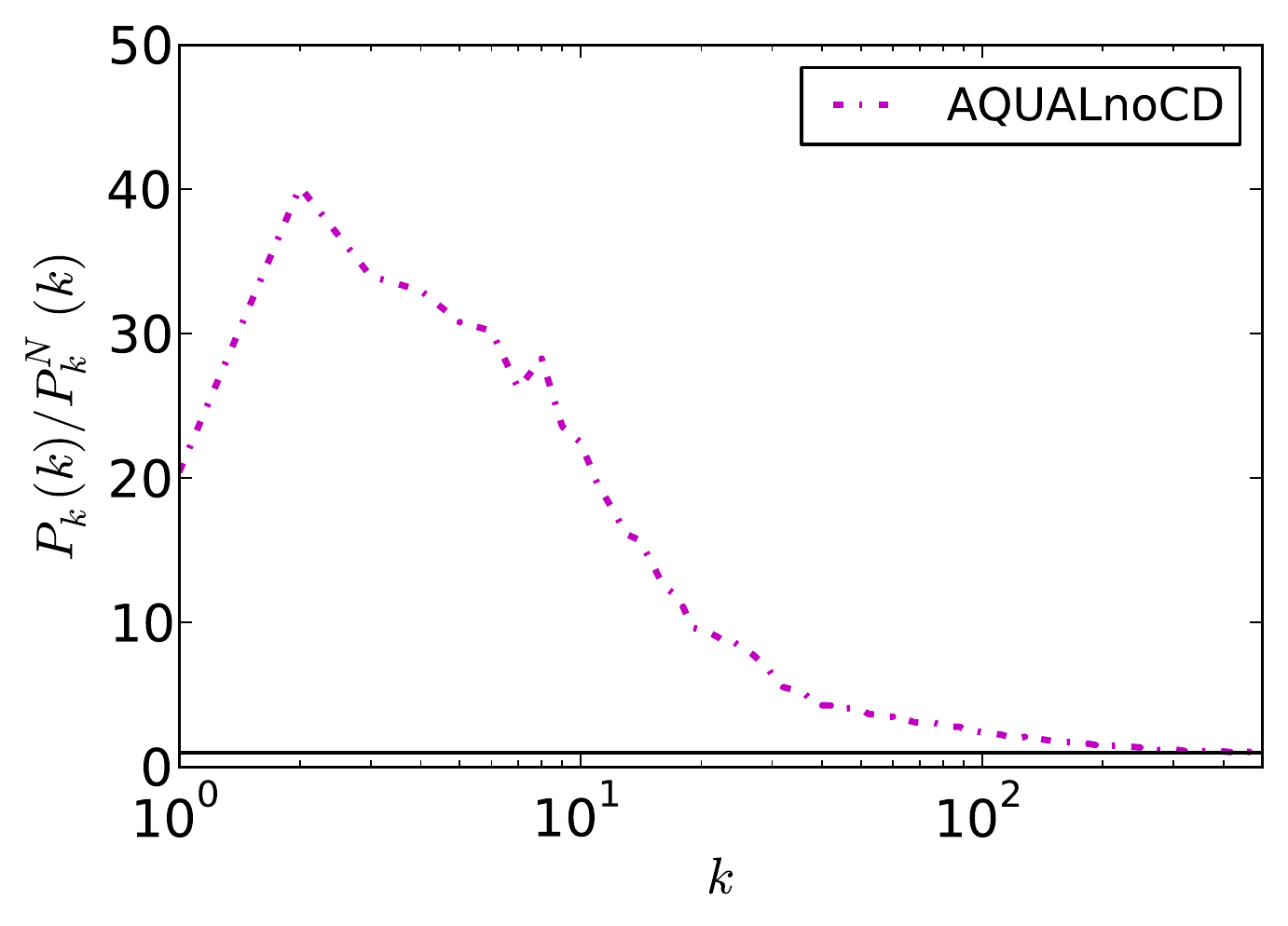}
\end{tabular}
\caption{Ratio of power spectra with respect to $\Lambda$CDM over time. Dot-dashed lines are at $a=0.3$, dashed lines are at $a=0.7$, and solid lines are at $a=1.0$. The AQUALnoCD model result is only plotted at $a=0.3$ due to the huge overproduction of structure already evident.}
\label{powspectime}
\end{figure*}

\begin{figure*}
\centering
\begin{tabular}{cc}
\includegraphics[width=7.0cm]{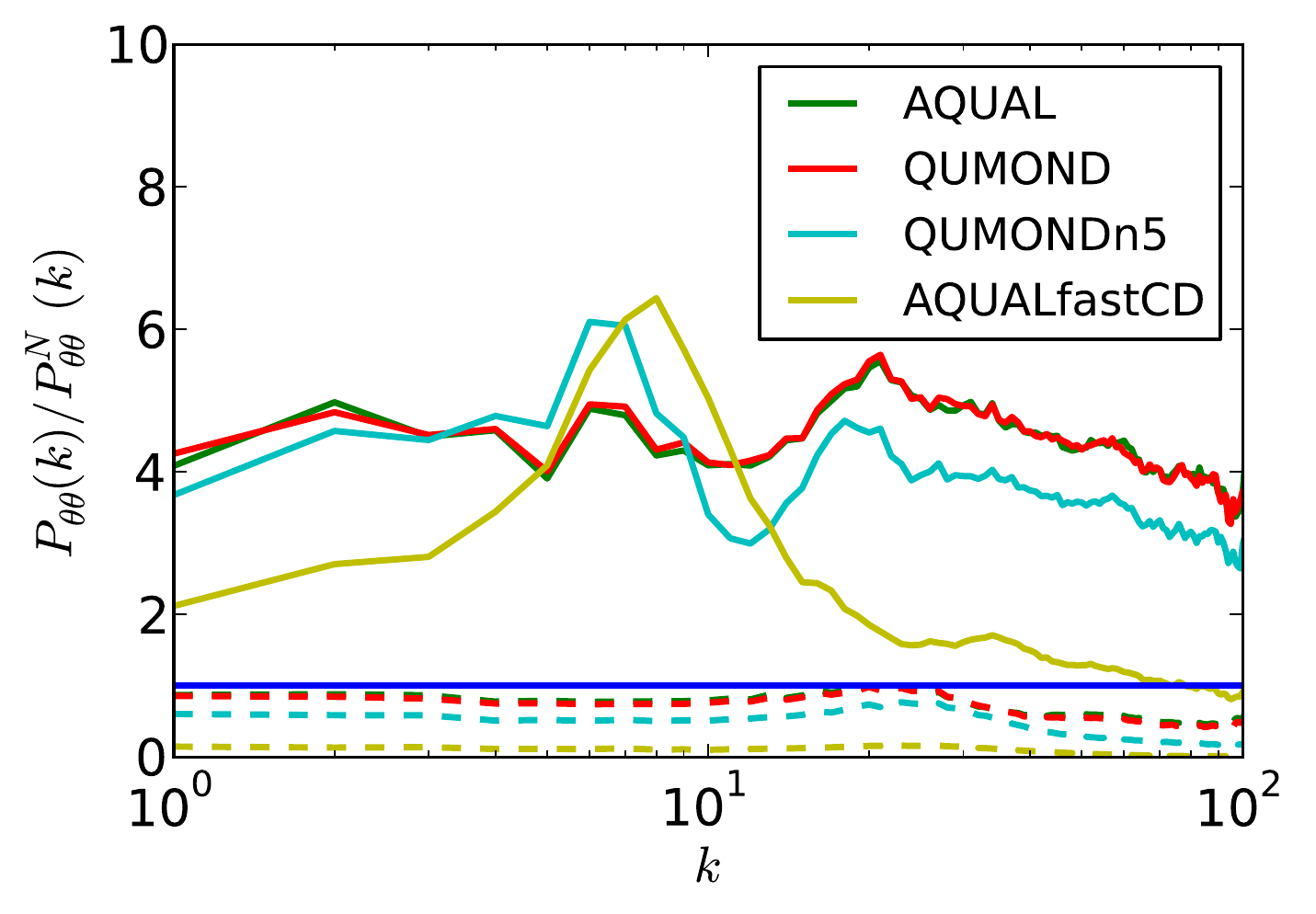} & \includegraphics[width=7.0cm]{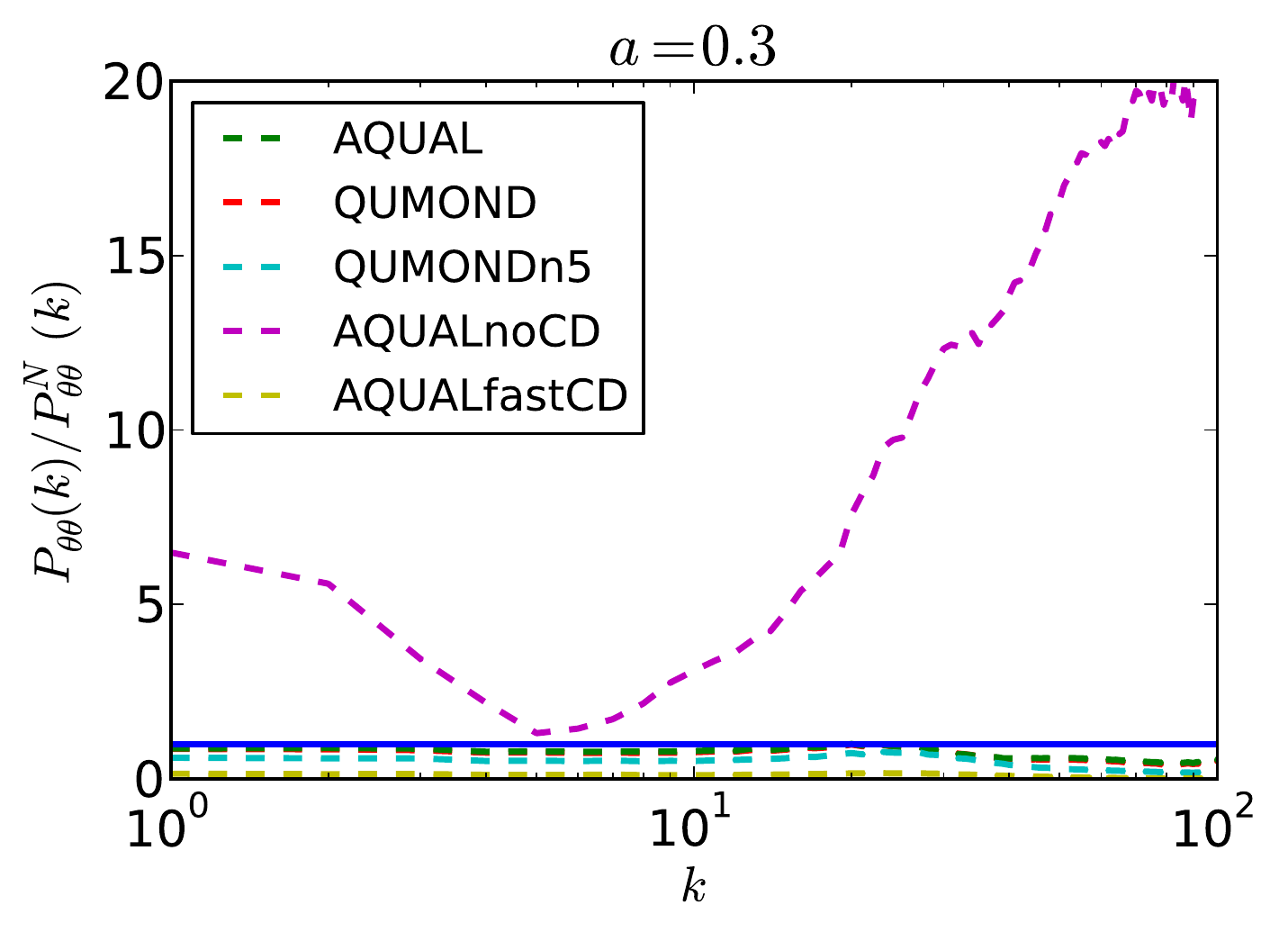}
\end{tabular}
\caption{Ratio of the velocity divergence power spectrum with respect to the $\Lambda$CDM result. The dashed lines are for $a=0.3$, and the solid lines for $a=1.0$. The thick blue line is at $P_{\theta \theta}/P_{\theta \theta}^N = 1$.}
\label{divpowspectime}
\end{figure*}

The time evolution of the ratio of the power spectra for the AQUAL and QUMOND simulations with respect to $\Lambda$CDM are given in the left panel of Fig.~\ref{powspectime}. We can now see clear evidence for that which is suggested in Fig.~\ref{densityplots}: structure formation proceeds more slowly at early times in the MOND simulations, becoming more rapid at later times, until finally catching up the Newtonian behaviour at late times. We can see that this is closely related to the strength of the MOND effect by suppressing the MOND enhancement, either through the more rapid transition function of QUMONDn5 or the faster cosmological transition of the MOND scale in AQUALfastCD, as shown in the centre panel of Fig.~\ref{powspectime}. Alternatively, a stronger modification of gravity at early times gives rise to an excessive amount of structure at late times. We can see this in the right panel of Fig.~\ref{powspectime} where the ratio of the power spectrum of the AQUALnoCD model (where the MOND effect is not suppressed at earlier times) to the $\Lambda$CDM model is shown, at the snapshot $a=0.3$. As this model has already produced far more structure than the $\Lambda$CDM run at $a=0.3$, it is not run beyond this epoch. Note that in all the MOND simulations, except AQUALfastCD, there is an overproduction of large scale structure relative to $\Lambda$CDM by $z=0$.

In the left panel of Fig.~\ref{divpowspectime} the time evolution of the ratio of the velocity divergence power spectrum with respect to the $\Lambda$CDM run is shown. At earlier times ($a=0.3$, indicated by dashed lines) we see that the Newtonian velocity divergence shows more power on all scales, as all the dashed lines lie below the thick blue line indicating $P_{\theta \theta}/P_{\theta \theta}^N = 1$. This difference is more pronounced at smaller scales. By $z=0$, however, all the MOND simulations (AQUAL, QUMOND, QUMONDn5 and AQUALfastCD) show a much larger velocity divergence relative to the standard model, on the order of $3-5$ times larger. Interestingly, the difference between the Newtonian run and the QUMONDn5 run is somewhat diminished at smaller scales, compared to the QUMOND and AQUAL runs. This is most likely due to the more rapid transition function pushing these scales closer to the Newtonian regime, while in the other MOND runs the slower transition function leaves these scales MONDian, so they undergo more velocity enhancement. The AQUALfastCD run shows an increased enhacement of the velocity divergence in intermediate scales, but less of an enhancement on all other scales. As the MOND effect is suppressed for a longer time in this model, the reduced enhancement at most scales is expected. The effect on intermediate scales appears to be due to these scales entering the MOND regime recently, where they have not formed substantial density gradients, and as such the velocity enhancement is strengthened when compared to the more developed structures (by $a=1$) in the other MOND simulations.

The velocity divergence power spectra for the AQUALnoCD and all other models at $a=0.3$ are given in the right panel of Fig.~\ref{divpowspectime}. We have already seen that all the other MOND models show less power at all scales in the velocity divergence at $a=0.3$ when compared to the standard model. The AQUALnoCD model at this epoch, however, already displays much higher velocity divergences at all scales, due to the far stronger MOND effect at earlier times. The dependence on scale of the velocity divergence enhancement of the AQUALnoCD model is essentially opposite to that of the AQUALfastCD model at late times shown in the right panel of Fig.~\ref{divpowspectime}. As the MOND effect can take hold much earlier in the AQUALnoCD run, structure has already formed, with the small scale velocity divergences already well beyond those of the $\Lambda$CDM model, where structure has not begun to form yet. On large scales, the weak gravitational accelerations leads to a strong MOND effect here as well, when compared to the standard model.

\begin{figure}
\centering
\includegraphics[width=7.0cm]{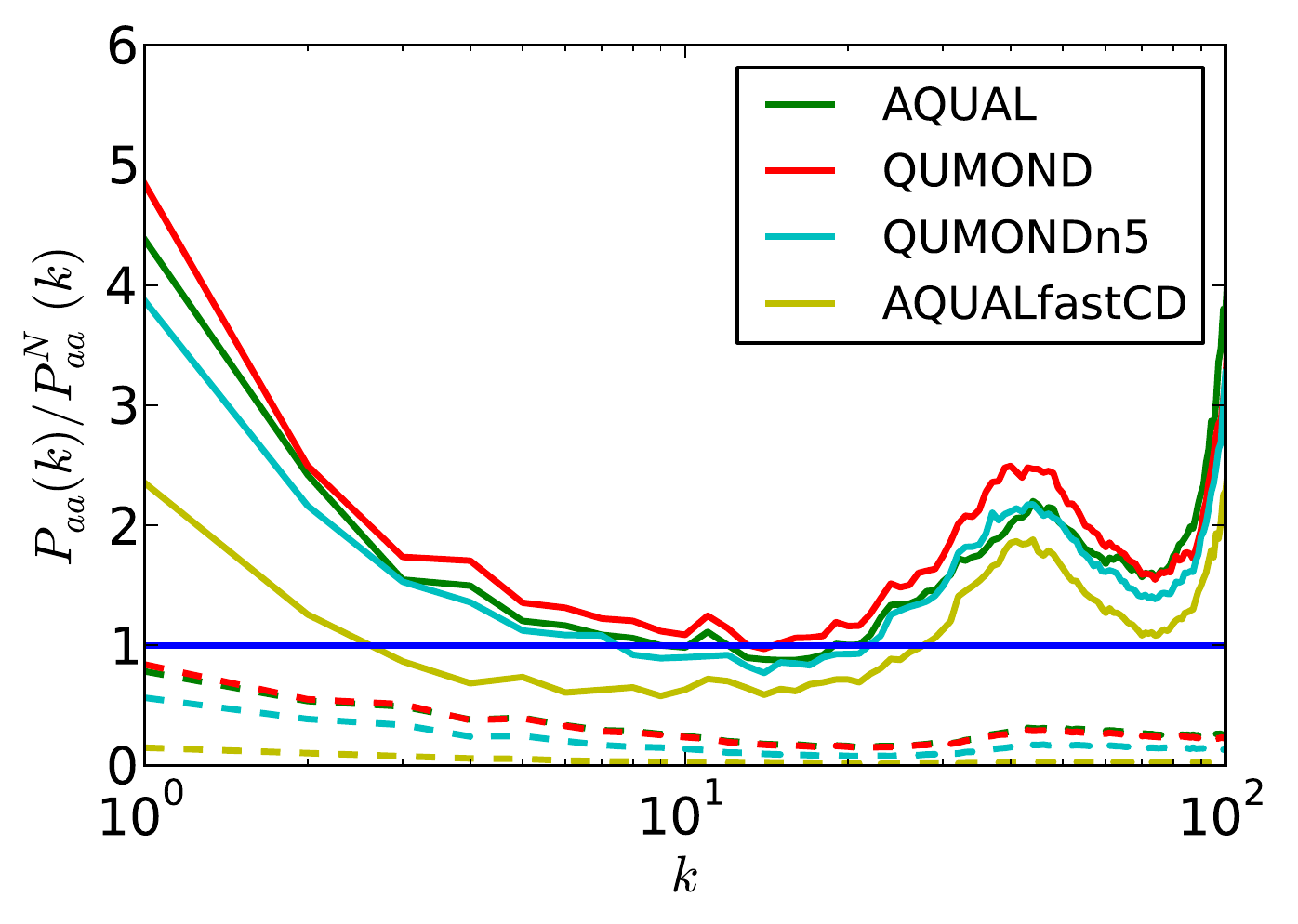}
\caption{Ratio of the acceleration power spectrum with respect to the $\Lambda$CDM result. The dashed lines are for $a=0.3$, and the solid lines for $a=1.0$. The thick blue line is at $P_{\theta \theta}/P_{\theta \theta}^N = 1$.}
\label{accelspec}
\end{figure}

The MOND effect on the velocity divergence is made more clear by calculating the power spectrum of the magnitude of the accelerations in the simulations. These are extracted to a regular grid using DTFE, as for the other quantities, and then Fourier transformed to find the power spectrum in $k$-space. The ratio of the AQUAL, QUMOND, QUMONDn5 and AQUALfastCD power spectra with respect to the $\Lambda$CDM model for $a=0.3$ and $a=1$ are shown in Fig.~\ref{accelspec}. The early time acceleration field is in fact below that of the standard model run, due to the much reduced matter content, and the early time suppression of the MOND effect. It is clear that by the end of the simulations, however, there is a significantly enhanced acceleration field at almost all scales. This is particularly clear on the largest scales, consistent with expectations from MOND \citep{nusser}. Interestingly, this acceleration enhancement reduces with increasing $k$ (decreasing length scale), until it is very similar to the Newtonian accelerations of the $\Lambda$CDM run for $k \sim 10$. At yet smaller length scales the MOND enhancement becomes stronger again, until there is another reduction at very small scales, before a final increase at the smallest scales that can be reliably probed in our simulations. The AQUALfastCD shows the expected suppression of the MOND effect as a vertical offset from the AQUAL and QUMOND power spectra. It is worth remarking, however, that even in this model the accelerations are somewhat enhanced both at large and small scales.

The power spectra in Fig.~\ref{accelspec} show the non-linearity inherent in MOND. At large scales, increasingly dominated by void regions, the accelerations are strongly increased as the Newtonian gravitational potential wells are weak. As the length scale decreases, there is a transition region where the MOND effect is sufficient to produce comparable accelerations for a Universe with a far lower matter content. At still smaller length scales, this leads to steeper potential wells than in the $\Lambda$CDM run: the increased density is associated to stronger Newtonian gravity as well, but the accelerations are still below the MOND scale and so are enhanced. The brief ``dip'' in the acceleration power spectra at $k \sim 80$ is likely due to the density contrasts becoming sufficiently high that the Newtonian accelerations are close to (or perhaps slightly above) the MOND scale, \emph{reducing} the MOND enhancement. There is also likely to be resolution effects at the very smallest scales, where the AMR grid size may be too large to properly account for the high potential gradients, leading to an artificial increase in the MOND accelerations.

\subsection{Average velocities as a function of density}
\begin{figure*}
\centering
\begin{tabular}{cc}
\includegraphics[width=8.0cm]{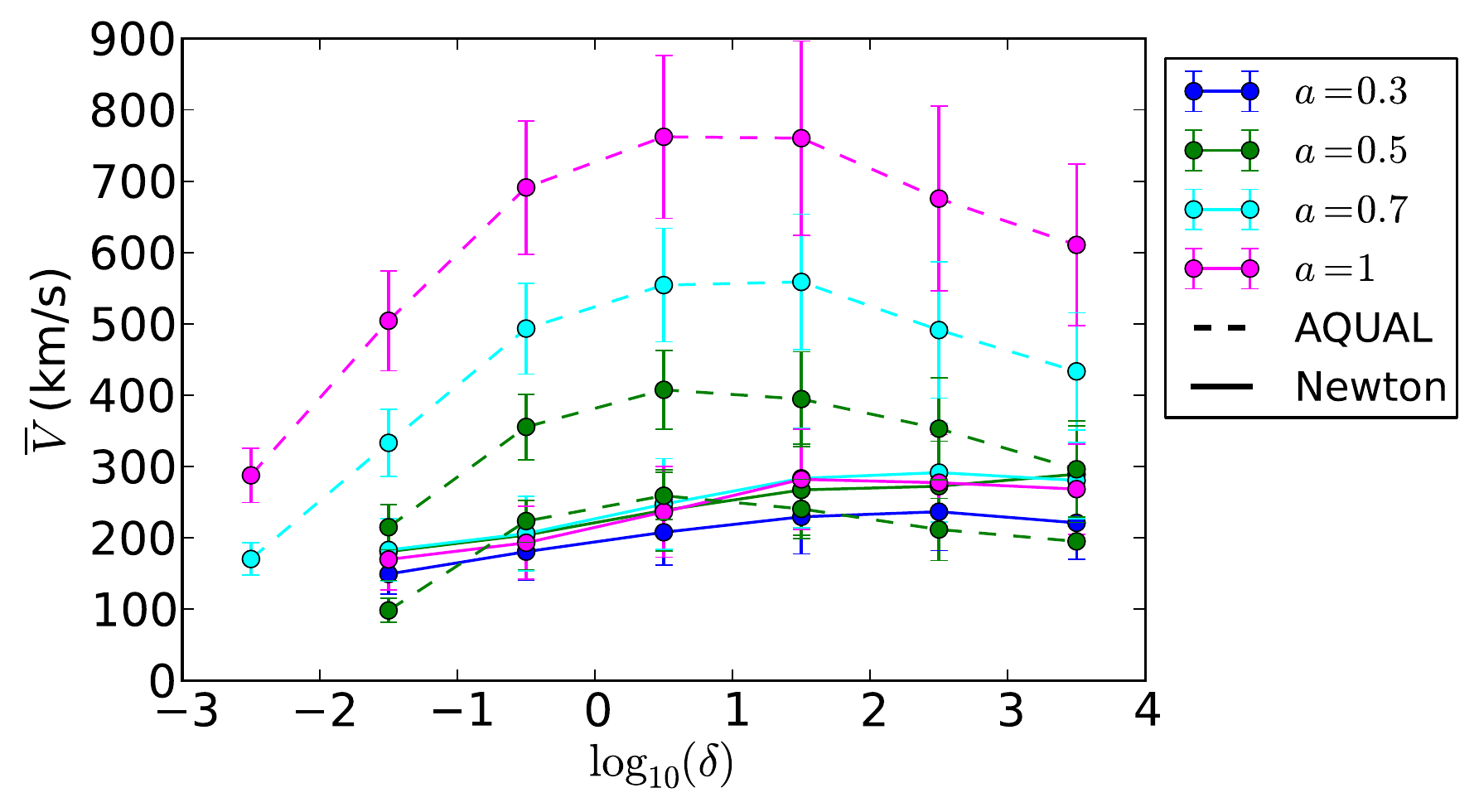} & \includegraphics[width=8.0cm]{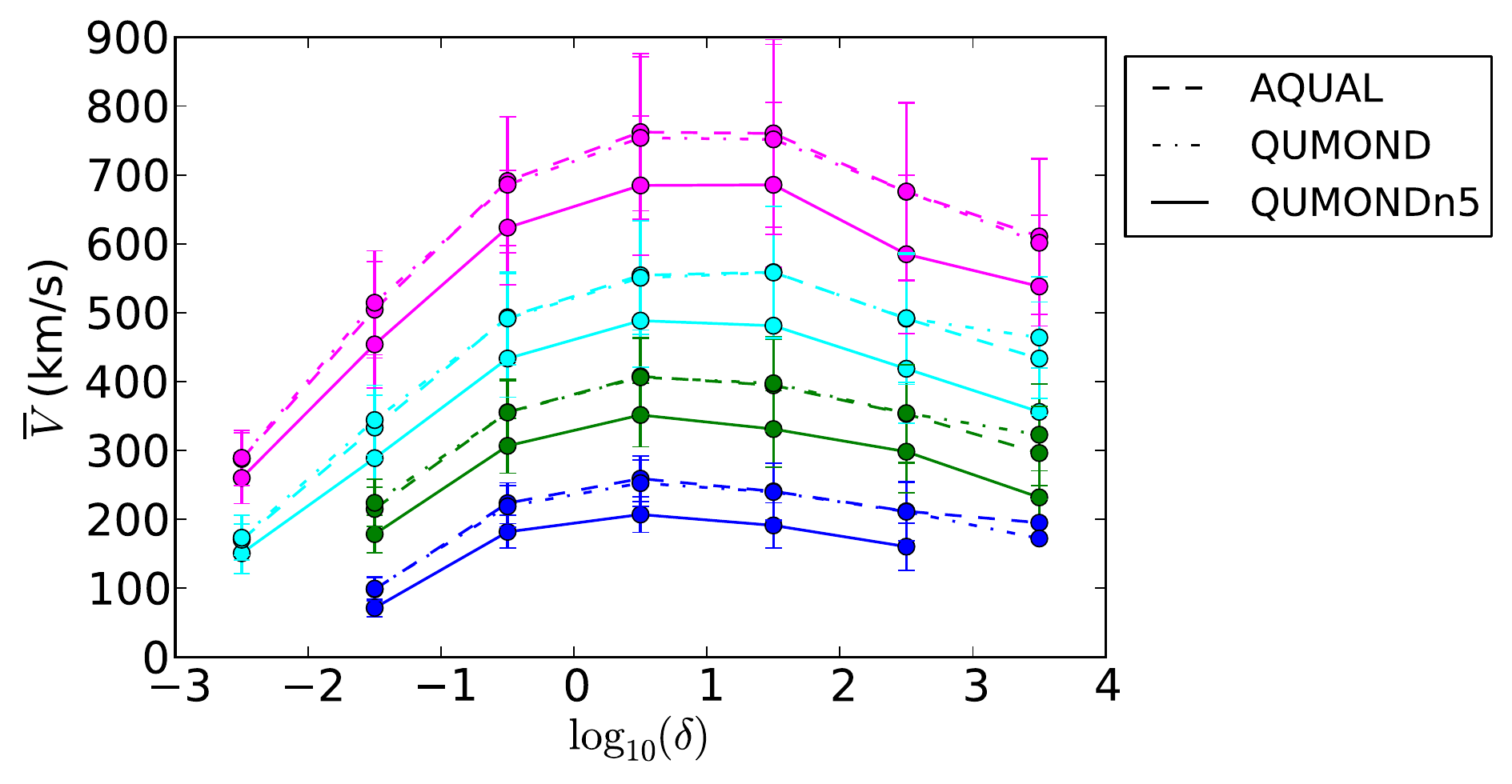}
\end{tabular}
\caption{Average velocity in density bins, derived from the DTFE density and velocity fields. The colours in the panel on the right correspond to the same scale factors as indicated for the panel on the left.}
\label{avgDensVel}
\end{figure*}

In order to determine if there is a density dependence in the MOND velocity enhancements, the average magnitude of the velocity field in bins of $\log(\delta)$ is calculated. Note that the MOND density field (extracted by DTFE) is used to calculate this relationship for the MOND velocities, and the $\Lambda$CDM density field is used for those velocities. This is shown in Fig.~\ref{avgDensVel}. While there is a notable amount of scatter (indicated by the $1\sigma$ error bars) there is a clear signal of a MOND enhancement of the velocities across all density bins, as we have already seen. The $\Lambda$CDM result shows remarkably little variation in the average velocities over time, and only a weak dependence on the density. This is consistent with previous studies of bulk flows in $\Lambda$CDM simulations, which may be in some tension with observations. Bulk flows will be investigated further in Section~\ref{bulk_velocities}.

Furthermore, we can see in Fig.~\ref{avgDensVel} that by late times the MOND velocity enhancement is clearly not homogeneous, with the most underdense and most overdense regions showing lower enhancements than the inner density bins. This indicates that the most pronounced velocity enhancements are not seen in the most underdense "voids" or in the most overdense "clusters", but are instead seen in the surrounding regions.

This is because in the most underdense regions the Newtonian gravitational potential will have very shallow gradients, and so any MOND enhancement of such tiny accelerations will still not lead to such a large velocity increase. In the case of overdensities, the gravitational accelerations will be increasingly close to or above the MOND scale for increasing density. This is consistent with other work showing that galaxy clusters are generally not within the MOND regime, and thus any missing mass problem there cannot easily be resolved by appealing to the MOND concept \citep{mondgalclusters,sterileneutrinos}.

The comparison with QUMONDn5 in the right panel of Fig.~\ref{avgDensVel} also shows that the rapid transition function leads to a suppressed MOND effect across all density bins, as evidenced by the offset between the AQUAL and QUMOND lines and the QUMONDn5 lines. This offset is, however, slightly smaller in the most underdense regions. In the limit of very low accelerations with respect to the MOND acceleration scale, the two forms of the MOND interpolation function ($n=1$ and $n=5$) begin to overlap, and so one would expect to see less difference between the QUMONDn5 run and the other MOND runs in the most underdense environments.

\subsection{Average accelerations as a function of density}
\begin{figure}
\centering
\includegraphics[width=7.0cm]{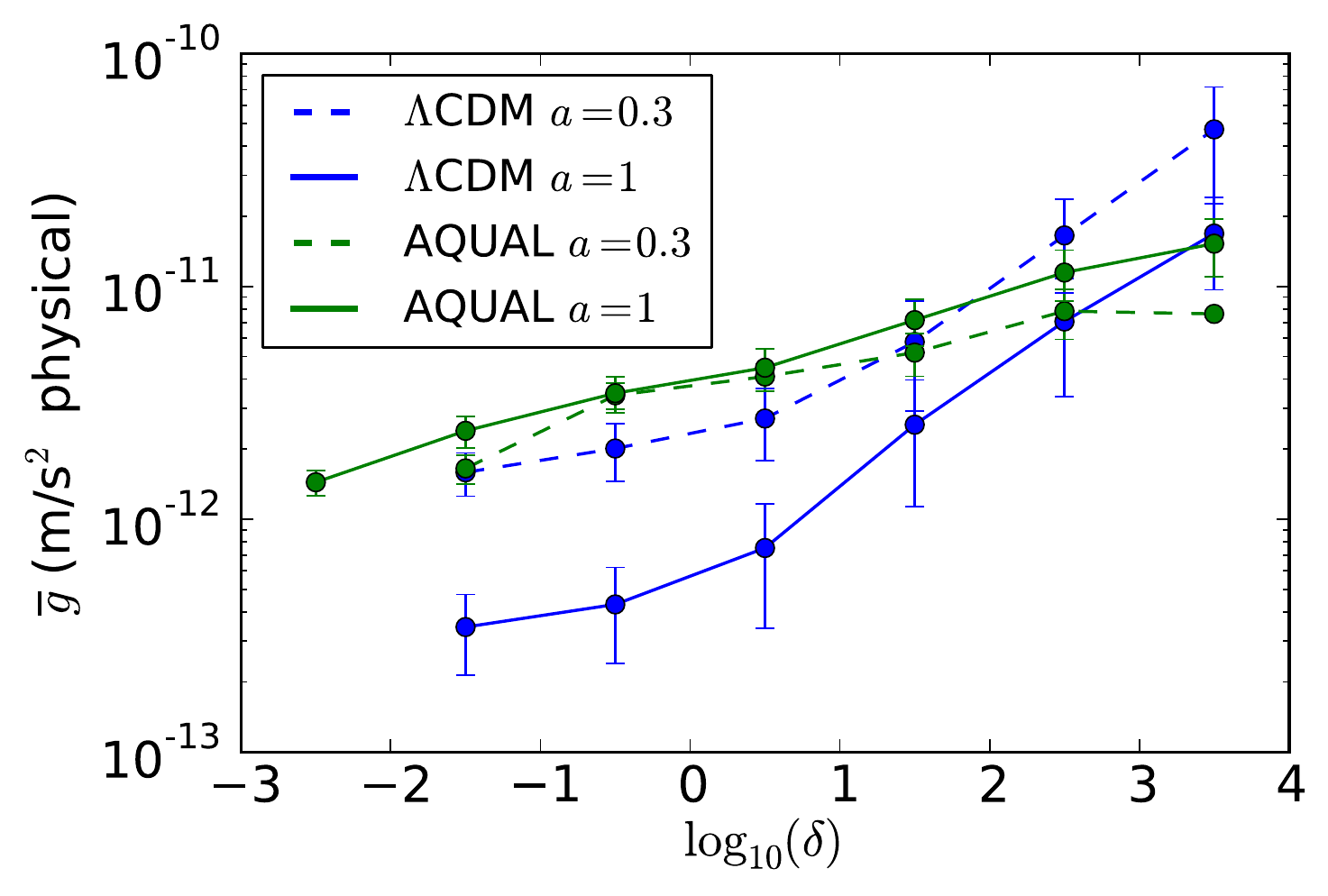}
\caption{Average acceleration in density bins (in physical units), derived from the DTFE density and acceleration fields.}
\label{avgDensAccel}
\end{figure}

As discussed earlier, at the end of Section~\ref{section:powspectra}, the non-linearity of the MOND effect leads to complex behaviour for the acceleration field as one transitions from underdense to overdense environments. Fig.~\ref{avgDensAccel} shows the average of the magnitude of the acceleration vector field, binned in density, as in Fig.~\ref{avgDensVel}. Note that we have converted the accelerations into physical units in this plot, and so the accelerations across all density bins \emph{reduce} for the $\Lambda$CDM run over time (the acceleration field decays as $\sim a^{-3}$ due to the expansion of the Universe). On the other hand, the AQUAL accelerations for $a=0.3$ and $a=1$ are of a similar order of magnitude in physical units, indicating the huge \emph{enhancement} of the acceleration field in this model.

One can clearly see that, at early times, the accelerations in the more overdense regions of the volume are higher in the $\Lambda$CDM run than the AQUAL run, again due to early suppression of the MOND effect and the reduced matter content. By $z=0$ the AQUAL accelerations in the underdense regions are far larger than in the $\Lambda$CDM run, but in the most overdense regions they are comparable, again illustrating the non-linear behaviour of MOND, and the lessening of the effect as the density increases. Nevertheless, it is clear that the accelerations, even in the highest density bin, are comparable with the MOND scale. The densities required to reach accelerations significantly beyond the MOND scale are undersampled in these low resolution simulations. Related to this, the commensurability of the high density accelerations at $z=0$ in Fig.~\ref{avgDensAccel} should not be interpreted as indicating that the AQUAL simulation is entering the Newtonian regime. The densities being considered are, of course, density contrasts with respect to the background density, which is some six times lower in the MOND simulations. Thus the MOND enhancement is, in general, still in effect, even in the high density regions, in order to produce a comparable magnitude of acceleration.

\subsection{Halo velocities}
The identification of ``halos'' in a MOND context may be considered as simply virialised overdensities of material within which one would expect galaxies to evolve. In this work the concept of a halo will be treated as a useful means of quantifying the late-time non-linear phase of the density distribution that arises from the initial conditions in the simulations. The conventional criterion for demarcating the spatial extent of a halo is the radius at which the overdensity is $200$ times that of the background density. This number derives from spherical collapse models applied to the standard $\Lambda$CDM paradigm, but is a somewhat arbitrary choice, given the cosmological dependence of this quantity, and the fact that the spherical collapse model is known to be a poor approximation to the true dynamics. A similar discussion regarding the ambiguities in the identification of ``halos'' in MOND simulations (using a sterile neutrino warm dark matter particle) may be found in \cite{katz_mond_vel}. In this Section the focus will be on the centre-of-mass velocities of the halos, which is not strongly affected by ambiguities in the halo definition. This quantity is effectively another way of probing the velocity field, this time treating overdensities as a coherent structure.

The AHF utility \citep{ahf} is applied to all of our simulations in order to find the halos. The rejection of halo particles based on their escape velocity is switched off, as this calculation assumes a Newtonian potential for the halo. For the Newtonian simulation a standard virial overdensity value of $\Delta = 200$ is used. This is adjusted for the MOND simulations to $\Delta = 35$, given that the MOND models contain roughly $6$ times less matter than the standard model. As stated earlier, however, the halo velocities are not strongly affected by this choice, as may be checked by varying the chosen overdensity value and comparing the resulting centre-of-mass velocity values. These are found to vary very little for choices of $\Delta$ between $20$ and $100$.

\begin{figure*}
\centering
\begin{tabular}{cc}
\includegraphics[width=7.0cm]{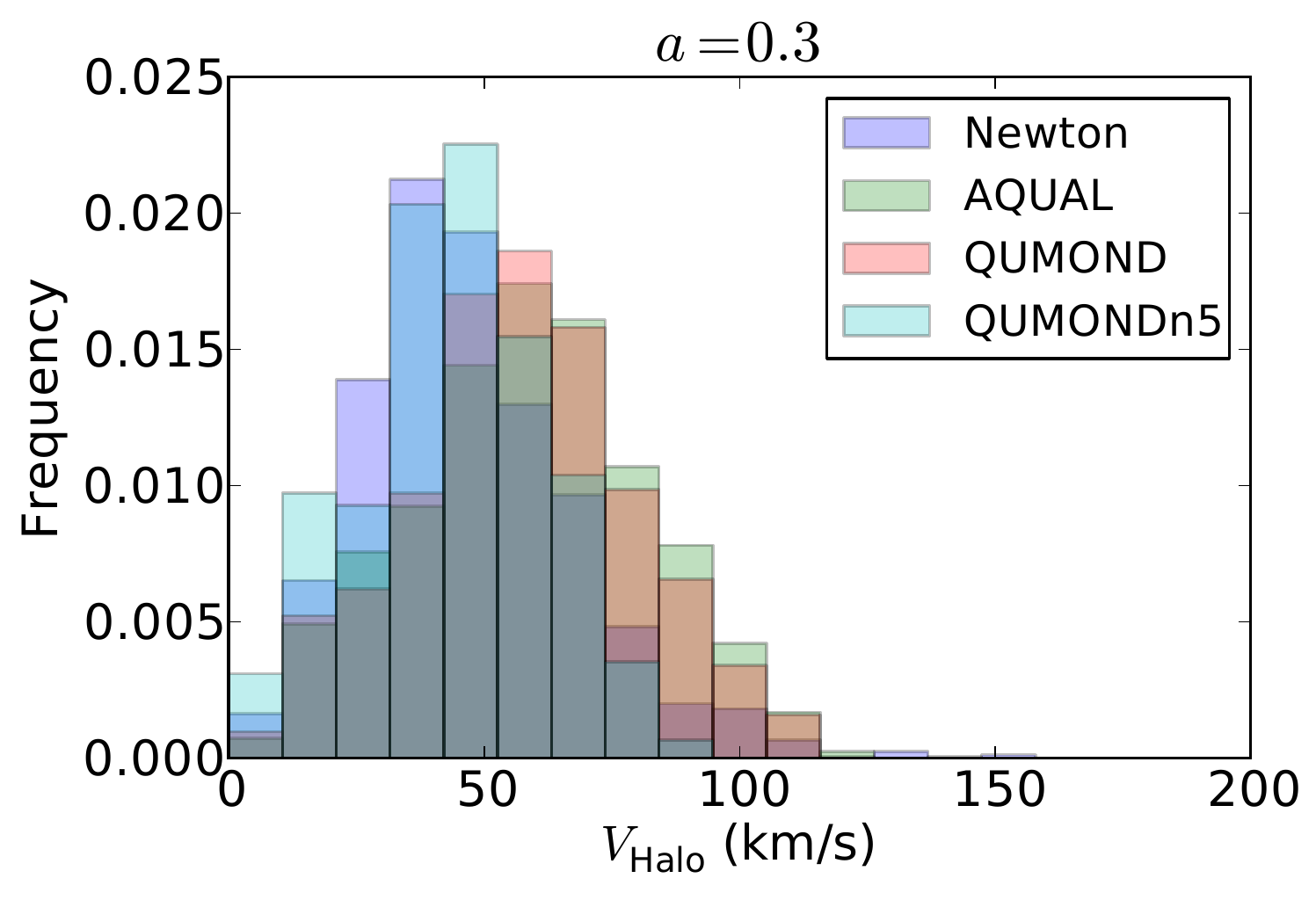} & \includegraphics[width=7.0cm]{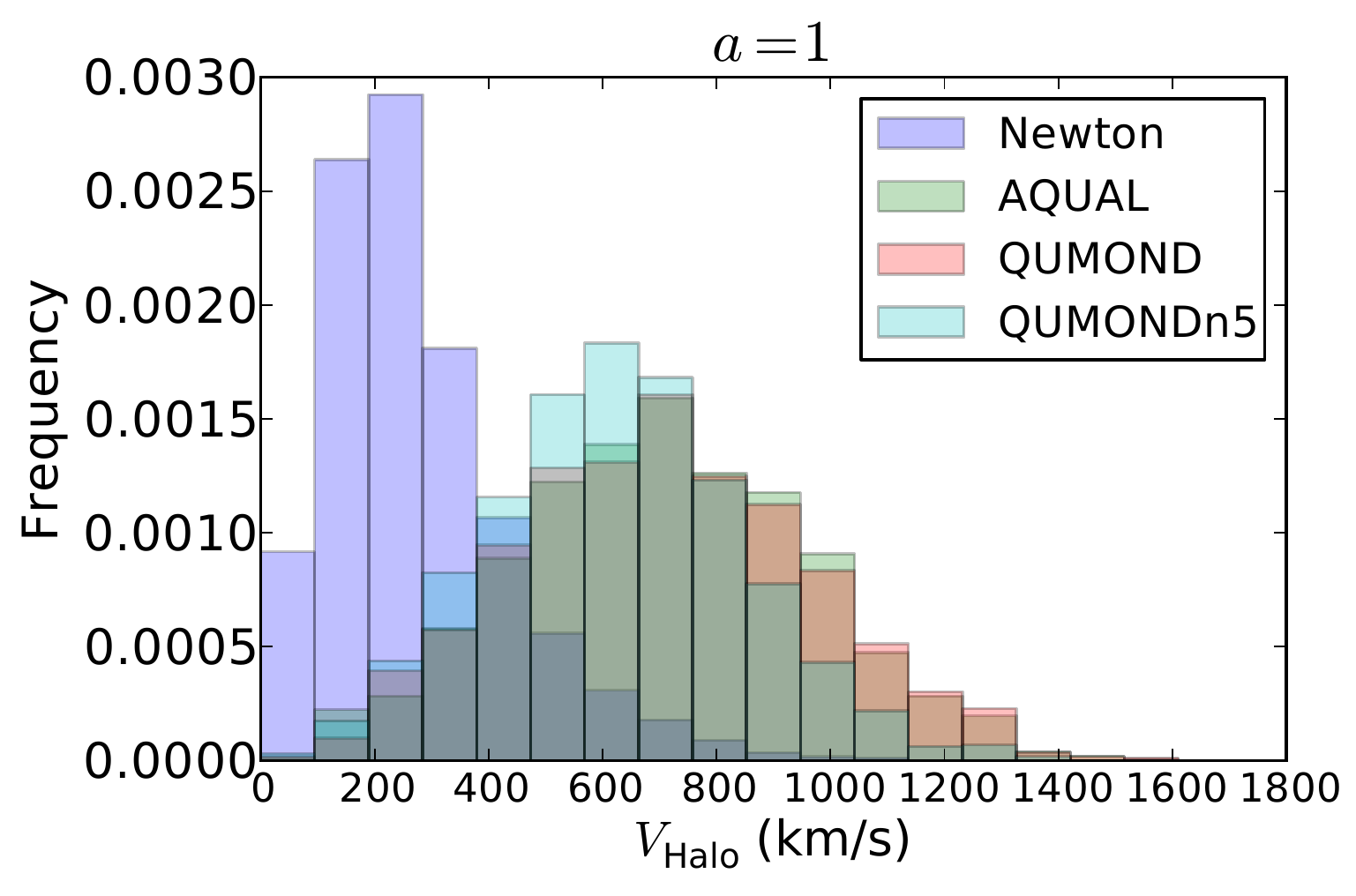}
\end{tabular}
\caption{Frequency distribution of halo CoM velocities. The left panel is for $a=0.3$, while the right panel shows the result at $a=1$.}
\label{haloVelocities}
\end{figure*}

The time development of the halo velocity distribution is shown in Fig.~\ref{haloVelocities}, with the $a=0.3$ and $a=1$ distributions shown. There is a clear velocity enhancement affecting the virialised structures, as already seen in the velocity fields. The spread of velocities in the MOND case evolves to become significantly larger than that of the $\Lambda$CDM simulation, with some halos reaching velocities in excess of $1200$~km/s. This is consistent with previous statements in the literature with respect to ``halo'' velocities in a MOND context. Previous studies have suggested that this may have consequences for interacting cluster systems such as the Bullet Cluster \citep{bulletoriginal1,bulletoriginal2}. This system is predicted to have a large relative collision velocity, possibly in excess of $2000$~km/s \citep{bulletsims1,bulletsims2}. Such a large collision velocity is considered to have a very low probability in a $\Lambda$CDM universe \citep{bulletlcdm1} (but see \citealp{bulletlcdm2}), whereas in a MOND context these velocities may well be more frequently attainable \citep{katz_mond_vel}. In the simulations of this study we see that the MOND velocities extend to approximately double those in the Newtonian case, and that low velocities, below $\sim 300$~km/s are much less common in the MOND simulations than in $\Lambda$CDM. These velocities are, however, much less than those reported in \cite{katz_mond_vel}. This point will be returned to in Section~\ref{bulk_velocities}.

\begin{figure*}
\centering
\begin{tabular}{cc}
\includegraphics[width=7.0cm]{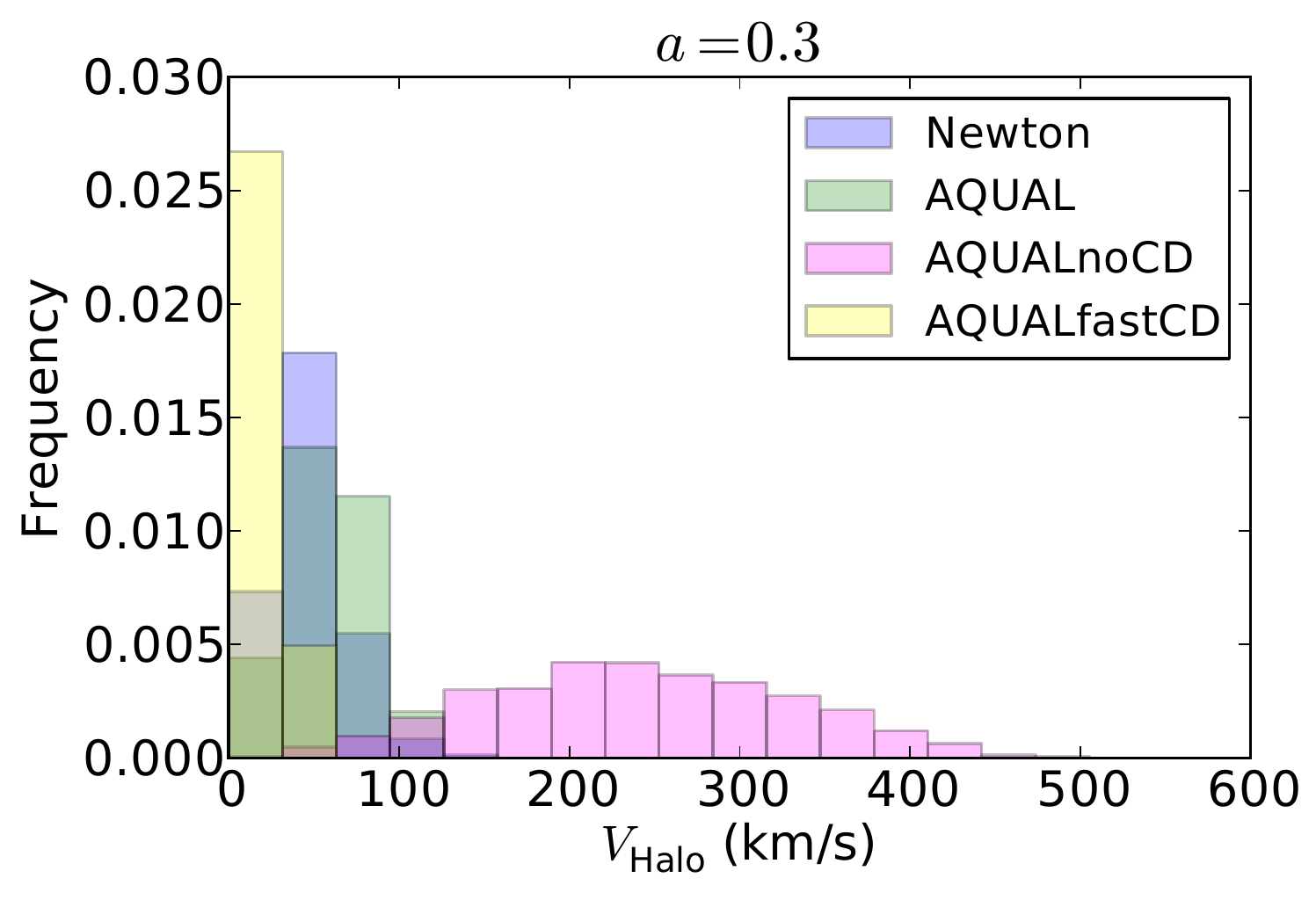} & \includegraphics[width=7.0cm]{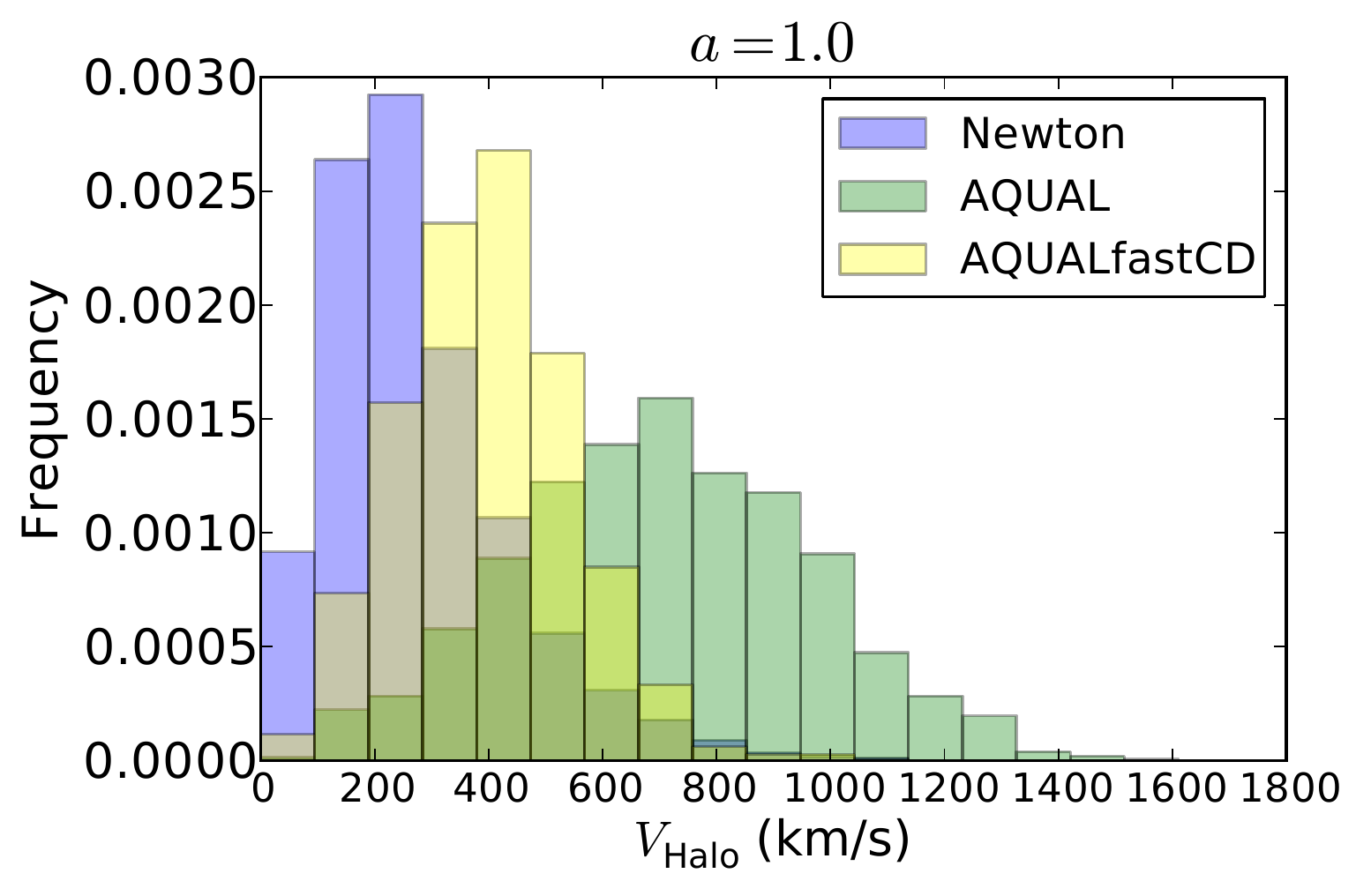}
\end{tabular}
\caption{Halo centre-of-mass velocity distributions, comparing the standard $\Lambda$CDM and AQUAL models with the NCD and FCD models.}
\label{haloVel_modCosmo}
\end{figure*}

In Fig.~\ref{haloVel_modCosmo} the early time halo velocity distributions of the AQUALnoCD and AQUALfastCD models, as compared with AQUAL and $\Lambda$CDM, are plotted in the left panel. The AQUALfastCD model halos have reduced velocities compared to both the standard AQUAL run and the $\Lambda$CDM run, while the AQUALnoCD model already has a large spread in velocities, with a median value of approximately $200$~km/s, far faster than any of the other models at this point in the cosmological evolution. The large tail of the distribution shows that some halos are already moving in excess of $400$~km/s. As the spread and mean of the distribution increases with time, the associated typical halo velocities are always higher in this model than in any other. At late times, given in the right panel of Fig.~\ref{haloVel_modCosmo}, the halo velocities in the AQUALfastCD model exceed those of the $\Lambda$CDM model, while being less spread and with a lower mean than for the standard AQUAL run. This again shows that the suppression of the MOND effect leads to a final cosmology that is in some sense ``in between'' the AQUAL and $\Lambda$CDM runs. It is worth pointing out, however, that the AQUALfastCD run still produces considerably higher halo velocities than the standard model run, even with significant suppression of the MOND effect until late times.

\subsubsection{Density dependence of halo velocities}
\begin{figure*}
\centering
\begin{tabular}{ccc}
\includegraphics[width=6.0cm]{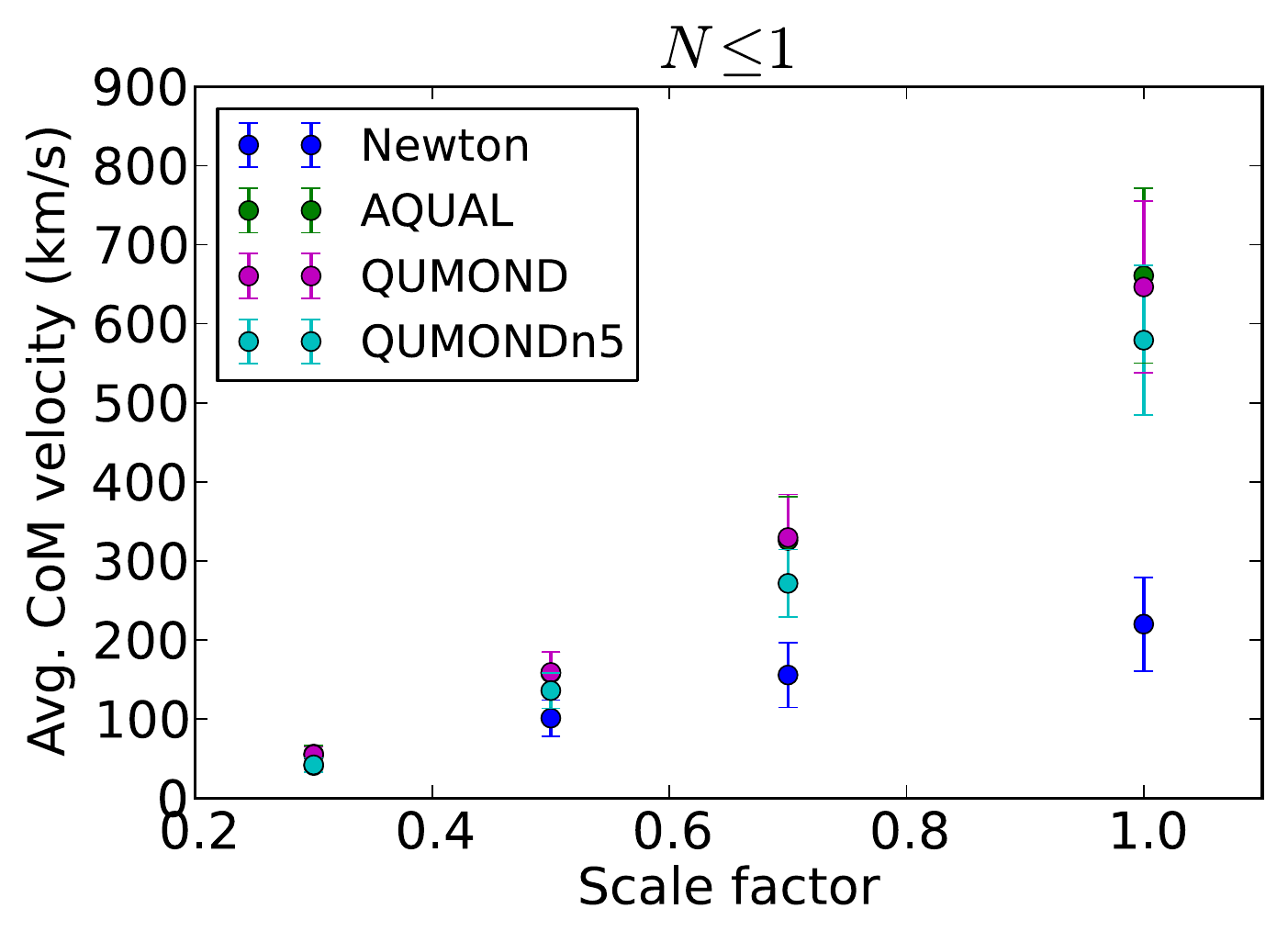} & \includegraphics[width=6.0cm]{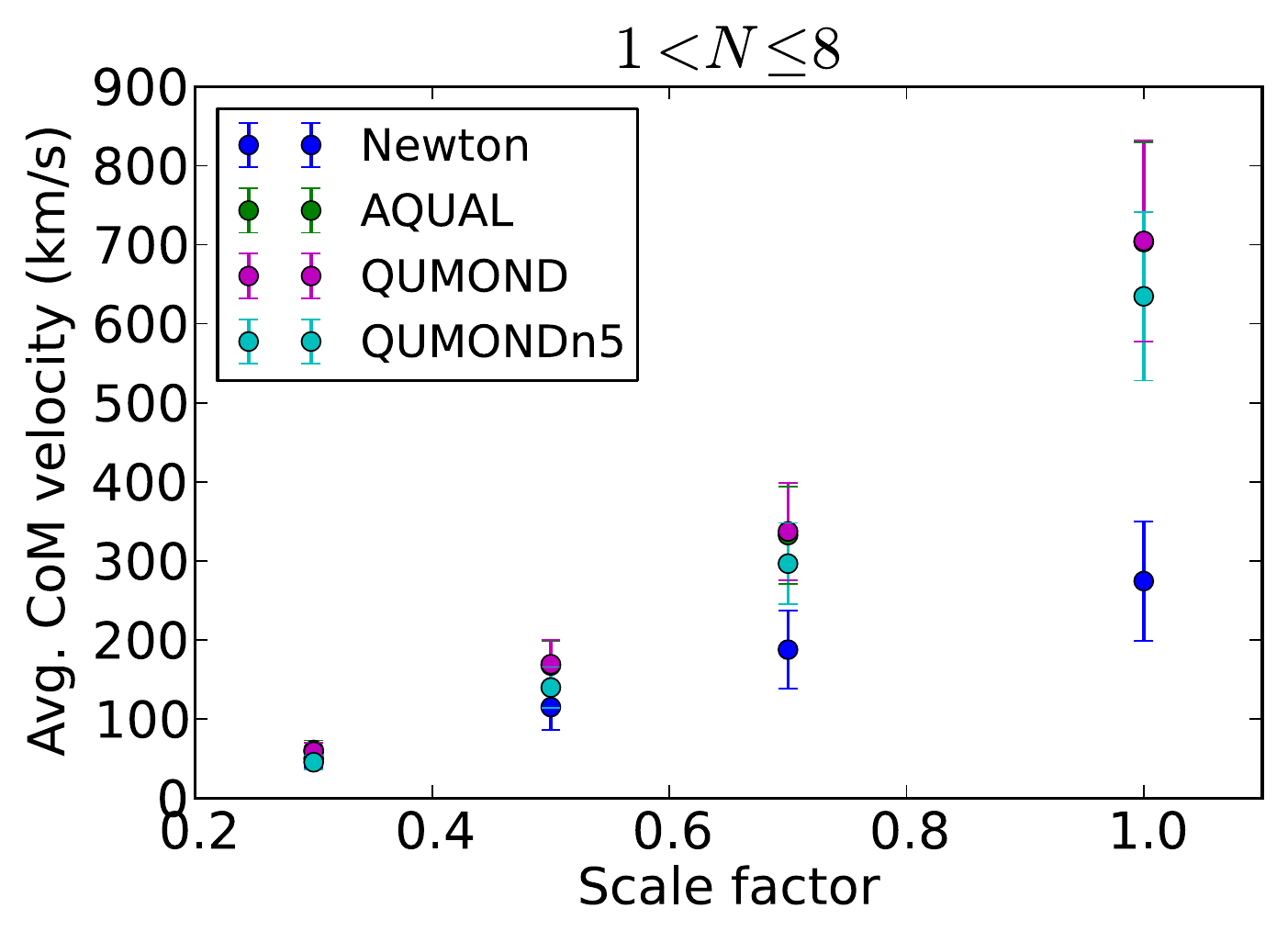} & \includegraphics[width=6.0cm]{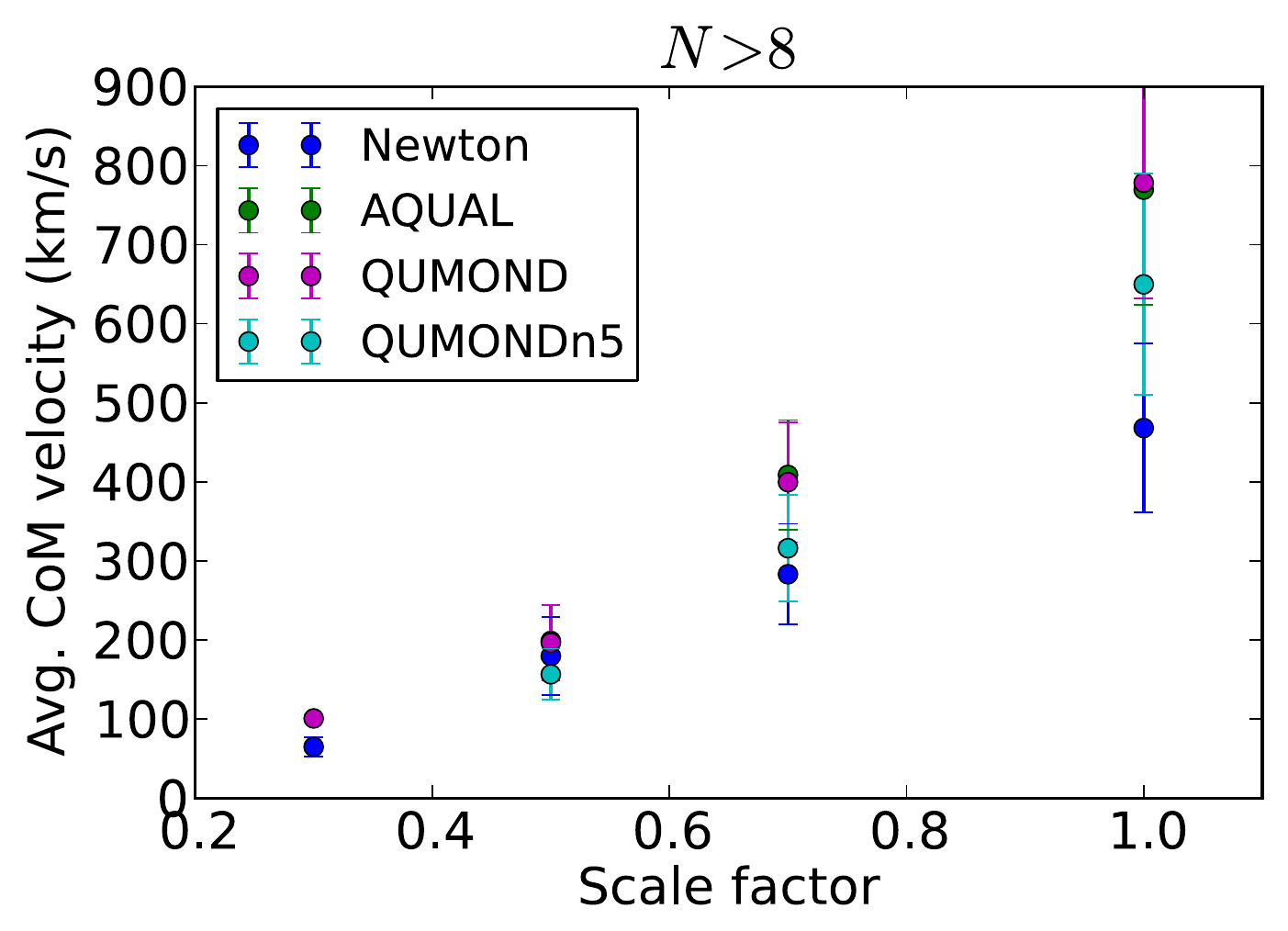}
\end{tabular}
\caption{Evolution of average halo centre-of-mass velocity, for halos in ``void'' environments ($N \leq 1$, where $N$ is the number of neighbours within $1 h^{-1}$~Mpc) in the left panel, ``sparse'' environments ($1 < N \leq 8$) in the central panel, and for ``dense'' environments ($N > 8$) in the right panel.}
\label{avgCoMdensesparse}
\end{figure*}

The left panel of Fig.~\ref{avgCoMdensesparse} shows the time evolution of the mean of the halo velocity distribution for those halos which have $1$ or less halos within $1 h^{-1}$~Mpc. The central panel of Fig.~\ref{avgCoMdensesparse} shows this information for those halos with $1 < N \leq 8$, where $N$ is the number of neighbours within $1 h^{-1}$~Mpc. The right panel of Fig.~\ref{avgCoMdensesparse} concerns halos with $N > 8$. These three criteria for halo neighbours will be taken to represent ``void,'' ``sparse,'' and ``dense'' environments. It is clear from these plots that the average centre-of-mass halo velocity becomes far higher in the MOND simulations than in the $\Lambda$CDM run, with a slight suppression of the difference in the case of a sharp MOND transition function, as in the QUMONDn5 run. Moreover, the difference in average velocities is much larger in the case of halos in the ``void'' and ``sparse'' environments. This is mostly due to the differences in the standard model, as the halo velocities are roughly twice as large in the ``dense'' environment by $z=0$, due to the halos falling into deeper gravitational potential wells. The halo velocities in the AQUAL and QUMOND simulations, however, show much less environmental dependence, with the $z=0$ average velocity difference between ``sparse'' and ``dense'' being only around $100$~km/s. 

Thinking of the MOND results in terms of velocity enhancements to the $\Lambda$CDM results, however, we can see that the MOND effect is \emph{stronger} in the ``void'' and ``sparse'' environments, consistent with the examination of the velocity fields earlier. In the most dense environments, the gravitational accelerations lead to higher velocities in both $\Lambda$CDM and MOND (still with some degree of enhancement in the MOND case), whereas in the environments we are labelling as ``void'' and ``sparse'' there is nevertheless sufficient structure to form halos, and it is there that we see a markedly more pronounced difference between the Newtonian and MONDian results. Environments with ``sparse'' halo distributions are precisely the cluster outskirts, filaments and void regions where we see the biggest enhancements in the DTFE-extracted particle velocity fields as shown in Fig.~\ref{avgDensVel}.

\begin{figure*}
\centering
\begin{tabular}{cc}
\includegraphics[width=7.0cm]{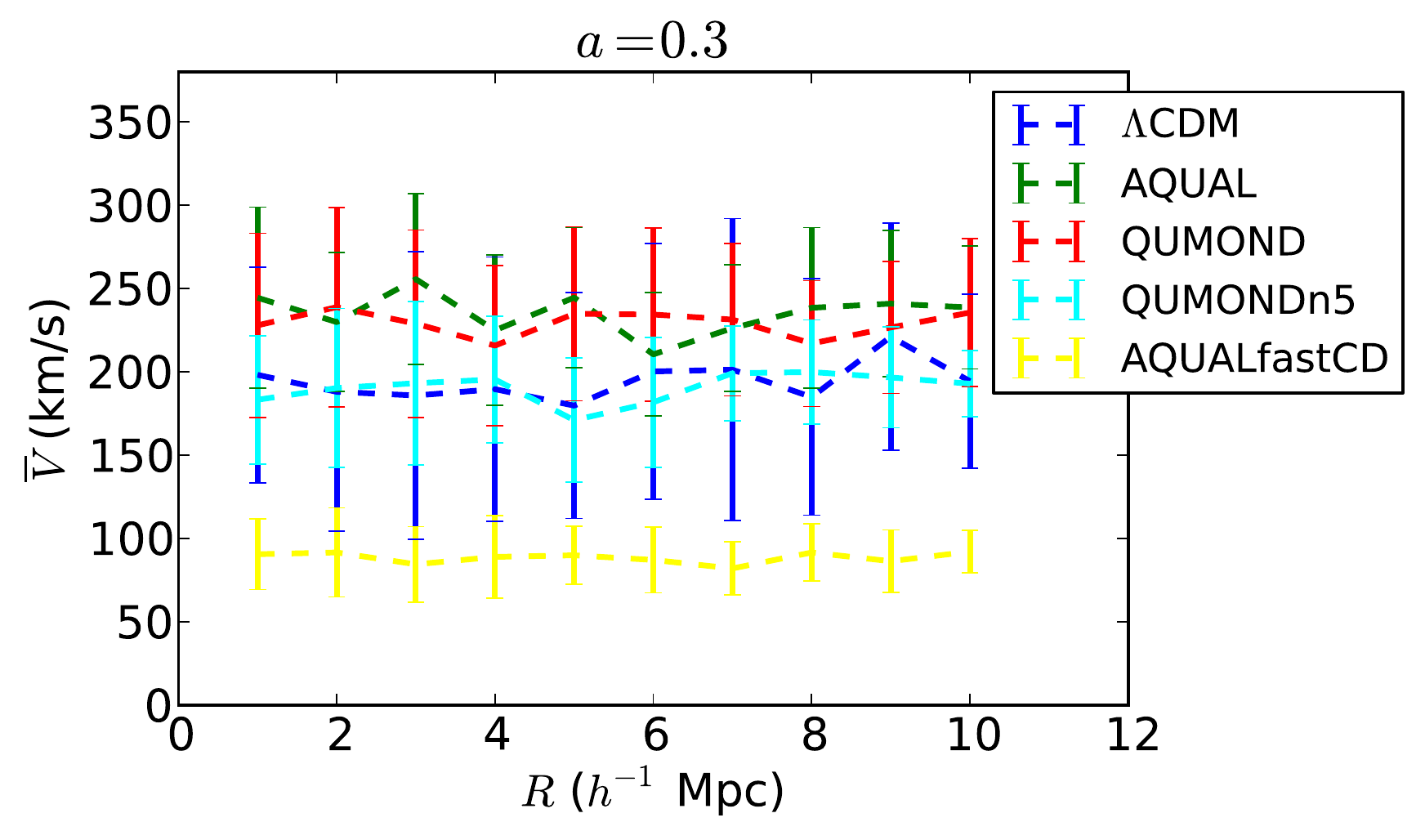} & \includegraphics[width=7.0cm]{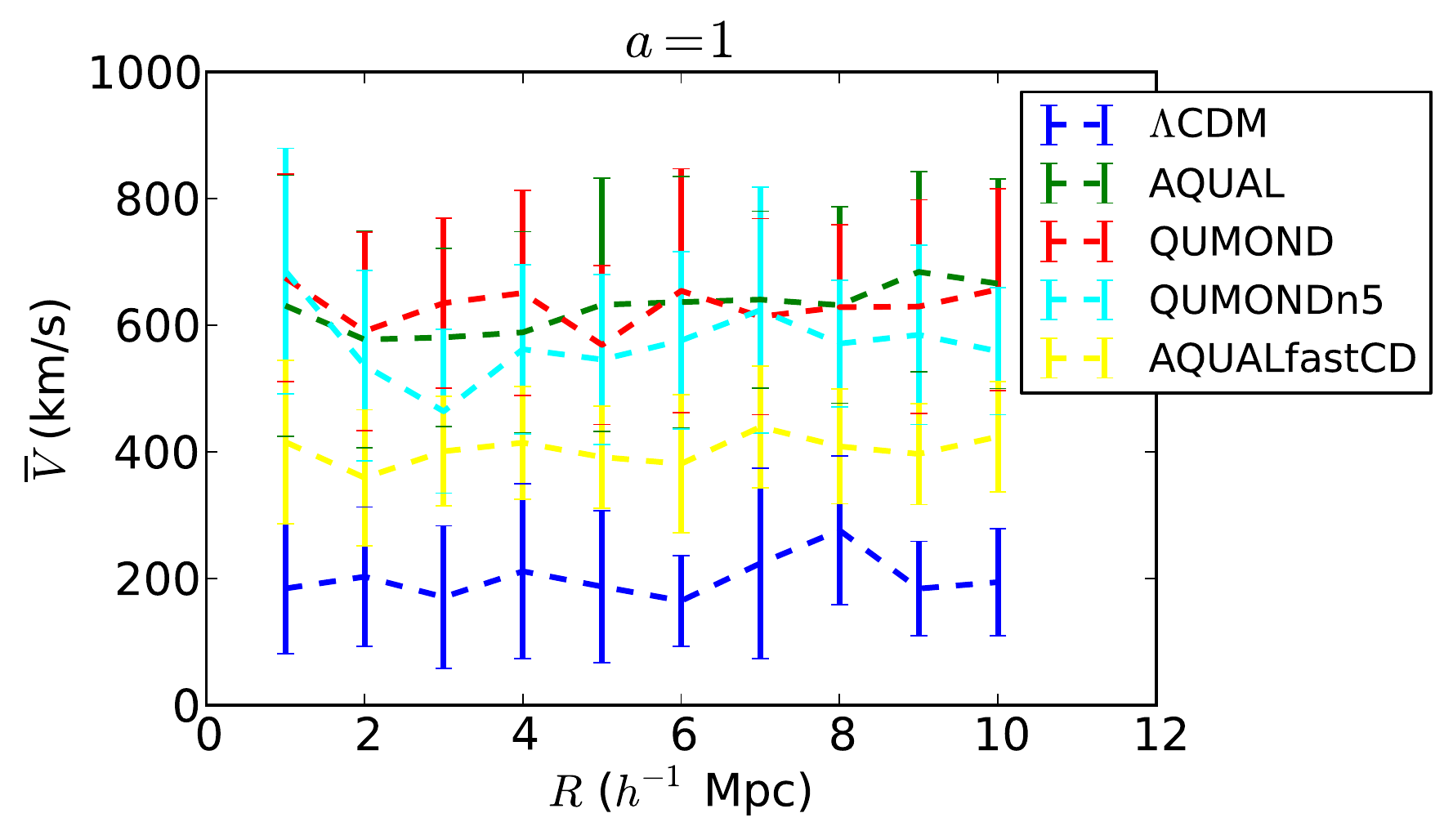}
\end{tabular}
\caption{Bulk velocities, calculated as the average velocity magnitude within $20$ randomly placed spheres of radii from $1$ to $10$~$h^{-1}$ Mpc. $\Lambda$CDM is compared with AQUAL, QUMOND, QUMONDn5 and AQUALfastCD at $a=0.3$ in the left hand panel, and at $a=1$ in the right hand panel. The error bars show the $1\sigma$ variation across the sample of randomly placed spheres.}
\label{bulkvel}
\end{figure*}

It is worth pointing out that the reduction in the velocity field seen in the highest density bin in Fig.~\ref{avgDensVel} is not reflected in the halo velocities in the ``dense'' environments. Of course, these quantities are very different: the highest density points in the DTFE-extracted grid correspond to the most virialised regions of the density field, i.e. to the particles that are forming the most massive halos, rather than the coherent motion of a virialised halo itself. The halo velocities are essentially a coarse-graining of the particle velocity field, and therefore cannot probe the highest density regions in the same manner as the underlying particles. Substantially increasing the resolution of these simulations, it is possible that low mass halos within a high density cluster would exhibit the reduced velocity enhancement shown for the high density bins in Fig.~\ref{avgDensVel}. The testing of this possibility is left for future work.

\subsection{Bulk velocities}
\label{bulk_velocities}
\cite{katz_mond_vel} addresses the possible observational impact of MOND on bulk velocities, claiming that MOND may be more consistent with observations of a large scale high-velocity ``dark flow'' (such as those of \citealp{kashlinsky_bulk,barandela}) than the $\Lambda$CDM model. It is also claimed that the high collision velocities of interacting clusters such as the Bullet Cluster may be more explainable in the context of MOND, rather than $\Lambda$CDM. In order to directly compare with observations, \cite{katz_mond_vel} derived the bulk velocities by extracting a cosmological evolution from the cluster velocities, and taking into account the apparent redshifts of clusters with respect to a chosen observer cluster in the simulation volume.

In this work, the bulk velocity (average magnitude of the velocity field) is calculated according to the prescription of \cite{jenkins}. While this is less directly comparable with observations, the main concern here is a comparison with the standard $\Lambda$CDM model, and the impact of modifications of the MOND behaviour.

Following \cite{jenkins}, the average magnitude of the velocity field (as sampled by the particles themselves, i.e. not extracted to the regular DTFE grid) in the simulation volume is calculated within spheres of different radii, where the sphere locations are randomly assigned. This gives rise to a systematic overestimate of the bulk velocities due to some of the spheres (particularly the larger ones) sampling close to the boundary of the simulation volume, where periodic boundary conditions are applied, giving rise to a repeated pattern of velocity measurements. This error applies to all the models of this study, however, and so is not of concern as long as no attempt is made to compare the simulation results to observations.

The measured bulk velocities in the $a=0.3$ and $a=1$ snapshots as a function of sphere radius are given in Fig.~\ref{bulkvel}. As shown before, the MOND modification of gravity leads to a significant enhancement of the bulk velocities at late times, with respect to $\Lambda$CDM. For the AQUALfastCD model the velocity enhancement is weaker, due to the MOND effect being suppressed for a longer time in the simulation. Nonetheless, there is an appreciable enhancement with respect to the standard model, with the bulk velocities being over $3$ times larger. At early times, the suppression of the MOND scale in AQUALfastCD maintains very low bulk velocities, some $45\%$ of the $\Lambda$CDM values. The QUMONDn5 model is comparable to the standard model, while the AQUAL and QUMOND runs are already showing higher bulk velocities than the standard model. This should be contrasted with the development of the density field, where the AQUAL and QUMOND runs lag behind $\Lambda$CDM, only catching up by late times. Note that the bulk velocities associated to the AQUALnoCD model at $a=0.3$ are not plotted, as the average across all sphere radii is $\sim 780$~km/s, approximately $4$ times higher than the standard model at this epoch.

\begin{figure}
\centering
\includegraphics[width=7.0cm]{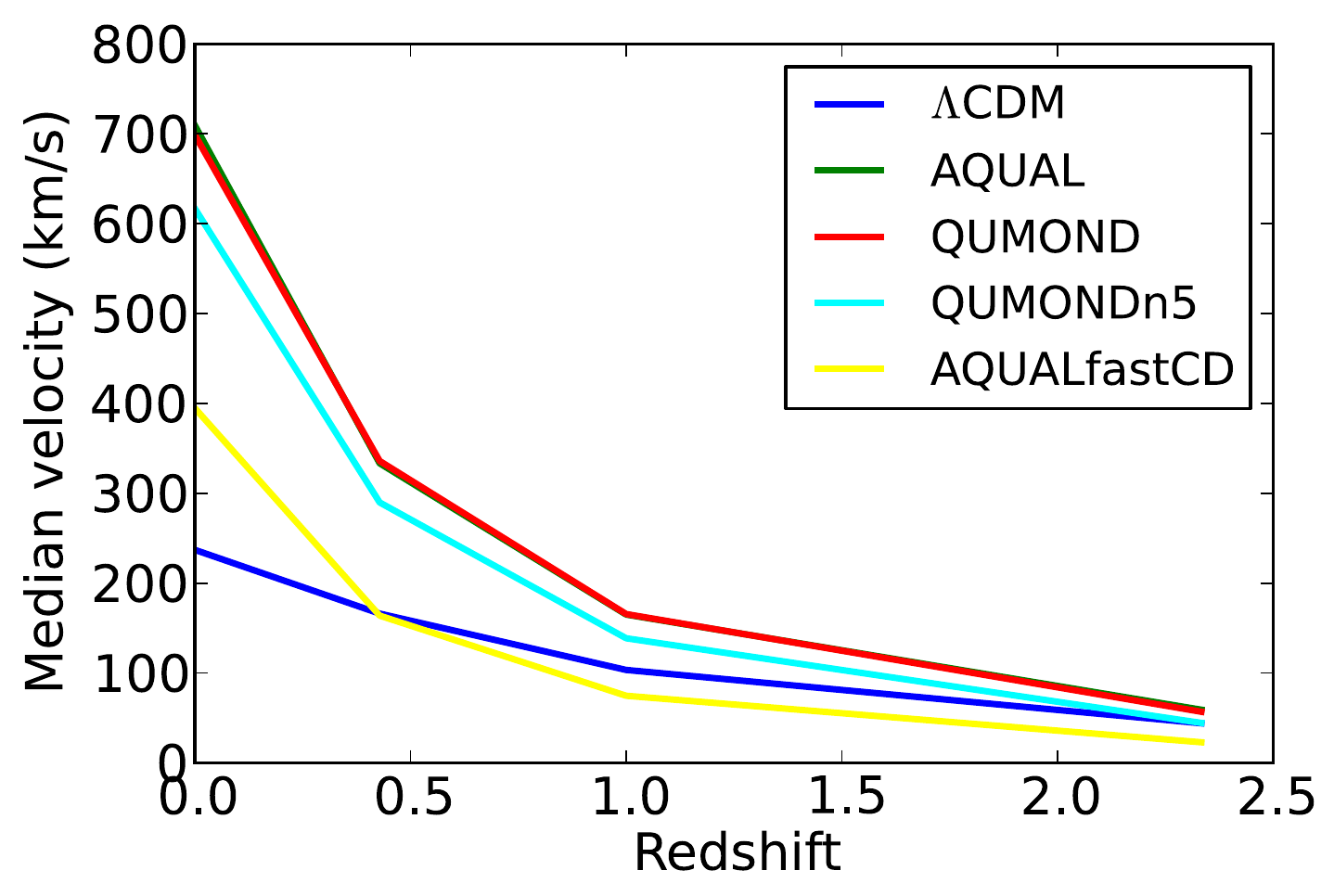}
\caption{The median velocities of halos at four different redshifts in our simulations, showing the faster growth of velocities in the MOND models.}
\label{velocity_function}
\end{figure}

These results also demonstrate that the bulk velocities in the MOND models change rapidly as the Universe evolves, increasing with time, in contrast to the far less pronounced evolution of the bulk velocities in the $\Lambda$CDM run. This is made more explicit in Fig.~\ref{velocity_function} where the median halo velocities at four different redshifts for the $\Lambda$CDM, AQUAL, QUMOND, QUMONDn5 and AQUALfastCD models are shown. It is clear that the MOND models all show a fast growth of the halo velocities as the Universe evolves, with suppression of this effect in the QUMONDn5 and AQUALfastCD models for the same reason as discussed earlier: the gravitational accelerations are typically less MONDian in these models, due to the more rapid transition function in the former case, and due to the suppression of the MOND scale in the latter.

These results may be compared with Fig. 4 of \cite{katz_mond_vel}. In that work, the median velocities for $z \leq 1$ are far higher than those shown here, in excess of $4000$~km/s. Two major differences between the present simulations and those of \cite{katz_mond_vel} are likely to be most responsible for the discrepancies in velocities. 

The first is the different initial conditions. The models of this study have a much reduced matter content in order to approximately match the power spectrum of the $\Lambda$CDM model at $z = 0$, whereas the models of \cite{katz_mond_vel} are instead based on initial conditions that provide a consistent background cosmology, as well as reproducing the CMB angular power spectrum. These requirements effectively lead to a direct replacement of cold dark matter with sterile neutrino dark matter, i.e. $\Omega_{\nu} = 0.2255$ in that work, somewhat similar to the value of $\Omega_{CDM} = 0.26$ for the $\Lambda$CDM model used here. The main advantage of a sufficiently light (or hot) sterile neutrino, however, is that the free-streaming length ensures there is very little clustering on galaxy scales, where pure MOND (without any additional unseen mass) is most effective. As reported in \cite{angus,angus2,katz_mond_vel} the problem of including this additional dark matter content is a significant overproduction of very massive halos. Furthermore, based on our results, it is highly likely that the much higher halo velocities seen in \cite{katz_mond_vel} are due to this increased matter content leading to steeper potential wells, which still undergo a MOND enhancement, and thus even faster accelerations that lead to much higher velocities.

The second major difference is the far smaller volume of our simulation: only $32^3 h^{-3}$~Mpc$^3$, compared to $512^3 h^{-3}$~Mpc$^3$ in \cite{katz_mond_vel}. Thus it is not possible to make a direct comparison of the bulk flow measurements of \cite{katz_mond_vel} as the box size used here is far smaller, leading to far less large-scale structure in the simulations. As the bulk flows appear to be dominated by larger structures, it is likely that higher velocities would be achieved in the present simulations when using a larger box size.

\subsubsection{Comparing with observations}
Observational studies of bulk flows and the velocity field in general do not yet present a consistent picture on all scales. Large scale measurements using the kinematic Sunyaev-Zel'dovich (kSZ) effect with the hot gas content of galaxy clusters have suggested very large scale bulk flows with amplitudes far in excess of that expected for $\Lambda$CDM \citep{kashlinsky_bulk,barandela}. Other studies, however, find contradictory results \citep{planck_pec_vel}. Although simulated bulk flow measurements in a larger cosmological volume may be larger than reported here, it is striking that the velocities in Fig.~\ref{bulkvel} are at the low end of the estimated bulk flow range reported in \citep{barandela} of around $600-1000$~km/s. It should be noted, however, that the length scale considered in \citep{kashlinsky_bulk,barandela} is $~300$h$^{-1}$Mpc, an order of magnitude larger than the largest scales probed in the present simulations.

A more recent study by \cite{feix}, using galaxy luminosities in the Sloan Digital Sky Survey (SDSS) \citep{sdss} to determine bulk flow measurements out to $z \sim 0.1$, found velocities fully consistent with $\Lambda$CDM. Similarly, the study by \cite{scrimgeour} using peculiar velocity data from the 6dF Galaxy Survey \citep{6dfgs} (probing scales of $50-70$h$^{-1}$Mpc) also found no inconsistency with the standard cosmological model. In addition, the use of Type Ia supernovae data to determine bulk flows \citep{feindt} also supports the $\Lambda $CDM model. These observations thus present a challenge to a MONDian cosmology of the form examined in this study. One may wish to appeal to a late-time ``switching on'' of the MOND effect to avoid excessive acceleration of structure inflows at early times. Despite the delayed onset of the MOND effect in the AQUALfastCD simulation, however, the bulk velocities are still approximately twice those of $\Lambda$CDM, and not consistent with the aforementioned measurements of bulk flows.

High velocity galaxy cluster collisions, such as the Bullet Cluster and Abell 520 \citep{trainwreck1,trainwreck2,trainwreck3}, also present a complex picture for both $\Lambda$CDM and MOND. While it is possible to accommodate high collision velocities more easily in MOND, as we have seen, it is unclear precisely how rare such configurations may be. The velocity enhancement in a MOND cosmology shifts the mean of the halo velocity distribution higher, implying more very high velocity halos in the tail. Thus Bullet Cluster-type collisions in MOND may not be as rare as in $\Lambda$CDM. It would be interesting to compare the statistical likelihood of such systems in the two models.

The Bullet Cluster, however, has traditionally been seen as problematic for MOND due to gravitational lensing maps that imply the existence of considerable amounts of invisible mass with a large offset from the visible baryonic material. Assuming a purely baryonic content for a MOND galaxy cluster, this would indeed be difficult to accommodate. If galaxy clusters in MOND also contain some additional unseen mass, however, it may yet be possible to reconcile Bullet Cluster-type systems with the MOND paradigm.

\section{Conclusions}
\label{conclusions}
The recently developed MONDification of the RAMSES simulation code, RAyMOND, has been applied to cosmology to examine the behaviour of non-linear structure formation in Modified Newtonian Dynamics, with particular emphasis on the large-scale velocity field. While these simulations are far from conclusive studies of MOND cosmology, they nonetheless suggest that the strongest differences between $\Lambda$CDM and MOND in structure formation studies are likely to be in observational reconstructions of the velocity field.

These results also conclusively demonstrate that the cosmological aspects of the RAMSES code are fully compatible with the modified gravity solver of RAyMOND, which succeeds in dealing with the peculiar nature of a cosmological simulation: very small density perturbations at early times with associated weak gravitational fields. This is a non-trivial test as such weak gravitational fields, having very small gradients, can lead to numerical difficulties in the non-linear AQUAL formulation. The results presented in this work show that the RAyMOND solver handles this well. Furthermore, the use of the non-linear Full Approximation Storage scheme for the Gauss-Siedel solver in RAyMOND is a well-known general technique applicable to many other modified gravity models, as well as models of dark energy coupled to dark matter/baryons. This is because the non-relativistic equation of motion of the dark energy scalar field is typically non-linear \citep{llinares}. Therefore, the modifications made to RAMSES to produce RAyMOND are presently being adapted to apply to coupled dark energy/dark matter models (Candlish et al., in prep).

Several major caveats apply to this work, with the first and foremost being the use of an unrealistic background cosmology (no dark energy) in order to use initial conditions with a substantially reduced matter content. In doing so, it is possible to reproduce the amount of structure (at least at smaller scales) produced in a standard $\Lambda$CDM simulation with the AQUAL and QUMOND runs. The fact that the initial conditions use a reduced matter content requires the removal of the effect of the cosmological constant from the models, and the use of the curvature component to roughly match the cosmological evolution of the scale factor to that of the standard model. While this is far from ideal, it is not currently possible to apply a known successful background MOND cosmology to the problem of structure formation. Therefore, this issue has been bypassed by assuming that some underlying relativistic MOND theory can indeed give rise to a sensible background cosmology, while at the same time the MOND effect plays a role in forming structures at late times. This study then focussed on the principal differences between the progression of MOND and $\Lambda$CDM structure formation.

Another important caveat is that no baryons have been included in these simulations. In a MOND context this is certainly a major omission, and it remains to be seen how the inclusion of baryons and baryonic processes affects the development of structure in such cosmologies and, even more significantly, the formation and evolution of galaxies.

Finally, although three possible forms for a cosmological dependence in the MOND scale have been studied, the chosen prescription for that dependence (Eq.~\ref{gamma}) is arbitrary, and may well differ substantially in a more complete MOND cosmology. It is also possible that time-dependent effects on non-cosmological timescales arise from a full MOND theory, modifying the nature of the MOND enhancement of the gravitational force, for example in galaxy collisions \citep{darkfluid}.

The fact that the AQUAL and QUMOND models lead to significantly larger velocities than the $\Lambda$CDM model, coupled with the results of the  AQUALfastCD model, where a suppressed MOND effect until late times still leads to enhanced velocities at $a=1$, leads to the central conclusion of this work: the use of MOND to form structure in a Universe with a reduced matter content requires the enhancement of the velocity field well beyond that seen in $\Lambda$CDM, potentially leading to a clear, unambiguous, observational test of MOND. In other words, if the large-scale measured velocity fields in the Universe can be unambiguously determined to be considerably higher than those of $\Lambda$CDM, this could constitute strong evidence in favour of some kind of modification of gravity in the manner of MOND, assuming that there is little room for major adjustment of the cosmological parameters of the standard model given constraints from other observables.

Given the present uncertainties regarding a complete MOND cosmology, there is a possibility that an alternative theory of MOND would produce a velocity field significantly different from that presented in this work. While such a possibility cannot be ruled out, the requirement of reproducing the successes of the standard MOND paradigm at galaxy scales is likely a stringent one for any future MOND model. If we assume that we do indeed live in a MOND Universe, then the precise transition between a fully relativistic MOND cosmology at early times, successfully reproducing observables such as the CMB power spectrum, to the present-day DM-free (at small scales) MOND Universe must occur at some point in the evolution of the Universe. The results of this study suggest that an enhanced velocity field is inevitable, unless this transition can be somehow engineered to occur at a very late time (although even the AQUALfastCD model showed significant velocity enhancement by $z=0$). Examinations of such possibilities beyond the standard MOND paradigm must remain the subject of future studies.

It is worth pointing out, of course, that there is a degeneracy here: a suppressed MOND effect may be counterbalanced by an increased initial matter content, as we have seen with the AQUALfastCD model being ``in between'' the AQUAL and $\Lambda$CDM runs. Of course, this is really a sliding scale, interpolating between the use of more matter (specifically dark matter) in order to produce the requisite structure with a weaker gravitational force, or using less matter and enhancing gravity. On the basis of this study, however, it appears to be extremely difficult to ``tune the dials'' in such a way that the matter content is reduced, with gravity enhanced through MOND, giving comparable present-day structure as the benchmark $\Lambda$CDM model, \emph{without} significantly enhancing the velocity field as well. The most extreme possibility, within the context of the standard MOND paradigm, would be to suppress the MOND effect throughout most of the cosmological evolution, and then to ensure its dominance at galaxy scales at very late times. This would require sufficient matter content to form structure in a non-enhanced gravitational field, with a suppression of structure at small (galaxy) scales so as not to spoil the late-time MOND behaviour.

These results are also consistent with the results of \cite{katz_mond_vel}. Although the velocities in the simulations studied here are likely lower than they would be given a larger cosmological volume, the very high velocities seen in the simulations of \cite{katz_mond_vel} using sterile neutrinos (constrained to match CMB observations and the background expansion history) are also closely linked to the increased matter content, which also leads to an excess of large scale structure. Furthermore, there are hints that even in the reduced matter content of the simulations employed here, there is some overproduction of structure on larger scales. It would be interesting to see if a modified cosmological dependence of the MOND acceleration scale, in conjunction with the suppression of small scale structure of the sterile neutrino DM model used in \cite{katz_mond_vel}, can help to resolve these issues. Thus the challenge of accounting for all the cosmological observables in a fully consistent MOND theory remains.

One very interesting aspect of the present investigation that deserves further study is the apparent dependence on the local density of the velocity enhancement effect. The results of this study, using both the velocity fields on the regular grids extracted by DTFE, and the halo analysis, suggest that the largest differences between $\Lambda$CDM and MOND velocity fields are likely to be found in the \emph{surroundings} of dense environments, i.e. the edges of voids, near filaments and possibly in galaxy cluster outskirts. To fully examine this possibility will require much higher resolution simulations than have been performed here. Furthermore, observational studies of the dynamics of structure formation in these regions, while challenging, are becoming increasingly possible thanks to modern observational capabilities. Comparisons between such observations and high resolution simulations in modified gravity will continue to provide a useful tool in testing the $\Lambda$CDM model.

\section*{Acknowledgements}
The author wishes to thank the anonymous referee for helpful comments that improved the paper. The author also wishes to thank Brad Gibson and Rory Smith for helpful comments and suggestions, and Alex Knebe for useful correspondence. The author gratefully acknowledges the support of FONDECYT grant 3130480 and from the Center for Astrophysics and Associated Technologies CATA (PFB 06).


\end{document}